\documentclass{article}
\pdfoutput=1

\usepackage{arxiv}
\usepackage[utf8]{inputenc} 
\usepackage[T1]{fontenc} 
\usepackage{hyperref} 
\hypersetup{
colorlinks = true,
linkcolor = black,
anchorcolor = black,
citecolor = black,
filecolor = black,
urlcolor = blue
}
\usepackage{booktabs} 
\usepackage{amsfonts} 
\usepackage{nicefrac} 
\usepackage{microtype} 
\usepackage{amsmath,bm}
\usepackage{graphicx}
\usepackage[flushleft]{threeparttable}
\usepackage{multirow}
\usepackage{natbib}
\usepackage{makecell}
\usepackage{caption}
\usepackage{subcaption} 
\usepackage{float}
\usepackage{placeins}
\usepackage{textcomp}

\newcommand{\figurehere}[1]{\begin{center}%
=========================\\%
Insert Figure #1 about here\\%
=========================\\%
\end{center}}
\newcommand{\tablehere}[1]{\begin{center}%
=========================\\%
Insert Table #1 about here\\%
=========================\\%
\end{center}}

\usepackage{array}
\newcommand{\PreserveBackslash}[1]{\let\temp=\\#1\let\\=\temp}
\newcolumntype{C}[1]{>{\PreserveBackslash\centering}p{#1}}
\newcolumntype{R}[1]{>{\PreserveBackslash\raggedleft}p{#1}}
\newcolumntype{L}[1]{>{\PreserveBackslash\raggedright}p{#1}}

\title{Estimating Knots and Their Association in Parallel Bilinear Spline Growth Curve Models in the Framework of Individual Measurement Occasions}

\author{
Jin Liu \thanks{CONTACT Jin Liu Email: Veronica.Liu0206@gmail.com, \textcircled{c}2021, American Psychological Association. This paper is not the copy of record and may not exactly replicate the final, authoritative version of the article. Please do not copy or cite without authors' permission. The final article will be available, upon publication, via its DOI: \url{10.1037/met0000309}}\\
Biometrics Department\\
Vertex Pharmaceuticals\\
 \And
Robert A. Perera\\
Department of Biostatistics\\
Virginia Commonwealth University \\
}

\begin{document}
\maketitle
\begin{abstract}
Latent growth curve models with spline functions are flexible and accessible statistical tools for investigating nonlinear change patterns that exhibit distinct phases of development in manifested variables. Among such models, the bilinear spline growth model (BLSGM) is the most straightforward and intuitive but useful. An existing study has demonstrated that the BLSGM allows the knot (or change-point), at which two linear segments join together, to be an additional growth factor other than the intercept and slopes so that researchers can estimate the knot and its variability in the framework of individual measurement occasions. However, developmental processes usually unfold in a joint development where two or more outcomes and their change patterns are correlated over time. As an extension of the existing BLSGM with an unknown knot, this study considers a parallel BLSGM (PBLSGM) for investigating multiple nonlinear growth processes and estimating the knot with its variability of each process as well as the knot-knot association in the framework of individual measurement occasions. We present the proposed model by simulation studies and a real-world data analysis. Our simulation studies demonstrate that the proposed PBLSGM generally estimate the parameters of interest unbiasedly, precisely and exhibit appropriate confidence interval coverage. An empirical example using longitudinal reading scores, mathematics scores, and science scores shows that the model can estimate the knot with its variance for each growth curve and the covariance between two knots. We also provide the corresponding code for the proposed model.
\end{abstract}

\keywords{Joint Development with Nonlinear Trajectories, Unknown Knot Locations,Individual Measurement Occasions, Simulation Studies}

\setcounter{secnumdepth}{3}
\section{Introduction}\label{intro}
Longitudinal analysis plays an essential role in various disciplines to investigate how the measurements of interest change over time. Researchers are interested in examining between-individual differences in within-individual change through analyzing such repeated measures.  The change patterns are likely to exhibit a nonconstant relationship to time to some extent if the study duration is long enough. According to \citet{Grimm2016growth}, the linear spline growth model (LSGM), which is also known as a piecewise linear latent growth model \citep{Chou2004PLGC, Harring2006nonlinear, Cudeck2010nonlinear, Kohli2011PLGC, Kohli2013PLGC1, Kohli2013PLGC2, Sterba2014individually, Kohli2015PLGC1} is one possible model to examine the individually nonlinear change pattern. With at least two attached linear pieces (see Figure \ref{fig:knot}), the linear spline (or piecewise linear) growth curve is capable of approximating more complex underlying change patterns. It has been widely employed in multiple areas, for example, the start of alcohol abuse \citep{Li2001PLGC}, the learning process of a specific task \citep{Cudeck2002PLGC}, mathematics ability development \citep{Kohli2015PLGC2}, reading ability development \citep{Kohli2017PLGC}, intellectual development \citep{Marcoulides2018PLGC}, and post-surgical recovery processes \citep{Dumenci2019knee, Riddle2015knee}.

\figurehere{1}

When analyzing a developmental process with a linear spline functional form, other than its initial status and rate of change of each piece, the change-points or `knots' at which the change rate has occurred must be determined. Driven by domain theories, empirical researchers may pre-specify knot locations \citep{Dumenci2019knee, Flora2008knot, Riddle2015knee, Sterba2014individually}. \citet{Marcoulides2018PLGC} also proposed a specification search by fitting a pool of candidate models and selecting the `best' one using the Bayesian information criterion (BIC). Additionally, \citet{Cudeck2002PLGC} fit a piecewise spline using multilevel modeling, showing that the knot can also be an estimated parameter, or even a random coefficient when it is not expected to be the same across all individuals. Knots have been estimated successfully using frequentist \citep{Cudeck2003knot_F, Harring2006nonlinear, Kwok2010simu, Kohli2011PLGC, Kohli2013PLGC1, Kohli2013PLGC2, Preacher2015repara, Liu2019knot, Liu2019BLSGM, Liu2019BLSGMM} and Bayesian \citep{Dominicus2008knot_B, McArdle2008knot_B, Wang2008knot_B, Muniz2011knot_B, Kohli2015PLGC1, Lock2018knot_B} mixed-effects models and growth models (including latent growth models and growth mixture models). 

The simplest linear spline function is a bilinear spline growth model (BLSGM, or a linear-linear piecewise model). This functional form helps identify a process that is theoretically two stages with different rates of change \citep{Dumenci2019knee, Riddle2015knee, Liu2019knot}; more importantly, it is also capable of approximating other nonlinear trajectories \citep{Kohli2015PLGC1, Kohli2015PLGC2, Kohli2017PLGC, Liu2019BLSGM}. \citet{Harring2006nonlinear} developed a BLSGM to estimate a fixed knot with the assumption that the knot is at an identical point in time for all individuals. They unified the functional form of the linear-linear change pattern through reparameterization by re-expressing a set of growth factors as linear combinations of their original forms. The model has been proven useful for examining development with two stages and estimating a fixed knot in a study of a procedural learning task \cite {Kohli2013PLGC2}. By relaxing the assumption of the same knot location across all individuals, \citet{Preacher2015repara} extended the BLSGM to estimate a knot while considering variability so that the knot is a growth factor (i.e., a random coefficient in the multilevel model framework) in addition to the intercept and two slopes. For interpretation purposes, \citet{Kohli2011PLGC}, \citet{Kohli2013PLGC1}, \citet{Grimm2016growth} and \citet{Liu2019knot}, \citet{Liu2019BLSGM} proposed to transform the mean vector and variance-covariance matrix of the reparameterized growth factors to the original setting, for the BLSGM with fixed and random knots, respectively.

\citet{Liu2019BLSGM} applied the BLSGM to estimate a knot with its variability for the developmental process in mathematics ability and found that the rate of development decreased following the knot and that the time for such a transition varied across individuals. However, developmental processes usually correlate with each other; accordingly, empirical researchers often desire to understand a joint development of multiple response variables of interest. For example, how the development of other skills, such as reading ability, correlates with the growth curve of mathematics scores.

One possible statistical method to analyze repeated measures of multiple constructs simultaneously is a parallel process and correlated growth model \citep{McArdle1988Multi}, also referred to as a multivariate growth model (MGM) \citep{Grimm2016growth2}. Earlier studies have shown that the correlated growth model is useful to analyze associated developmental processes where trajectories can be either linear or nonlinear. For example, \citet{Robitaille2012PLGM} detected a significant intercept-intercept and slope-slope association when analyzing a joint development of visuospatial ability and processing speed by employing the MGM with linear growth curves. \citet{Blozis2008MGM} demonstrated how to model parallel growth in two continuous response variables with nonlinear change patterns using parametric functions, such as polynomial and exponential, proposed by \citet{Blozis2004MGM} using \textit{LISREL}. Additionally, \citet{Ferrer2003MGM} suggested alternative models, such as latent difference scores dynamic models that can also be applied to analyze bivariate nonlinear developmental processes. However, to our knowledge, no existing studies evaluate a joint development considering piecewise change patterns in the structural equation modeling (SEM) framework. In this work, we propose parallel BLSGM (PBLSGM) by extending the BLSGM \citep{Liu2019BLSGM}, which estimates a knot with its variance, to the multivariate growth model framework. With the PBLSGM, it is of interest to assess the knot-knot association other than the intercept-intercept and slope-slope association between at least two repeated outcomes. We also extend the (inverse-) transformation functions and matrices that transform growth factors in two parameter-spaces developed in \citet{Liu2019BLSGM} for the PBLSGM so that its estimates can be directly interpreted.

Similar to \citet{Liu2019BLSGM}, we propose the PBLSGM in the framework of individual measurement occasions due to possible heterogeneity in measurement times in a longitudinal study \citep{Cook1983ITP, Finkel2003cognitive, Mehta2000people}. This can occur when participants differ in age at each measurement occasion in developmental and aging studies where the response is more sensitive to the change in age than that in the measurement time. Another possible scenario of heterogeneity in measurement occasions may result from when longitudinal responses are self-initiated. For example, in an adolescent smoking study, longitudinal records were collected from questionnaires that were asked to complete immediately after smoking \citep{Hedeker2006long}. With a definition variable approach, we fit the proposed PBLSGM with individual measurement occasions. \citet{Mehta2000people} and \citet{Mehta2005people} termed the `definition variables' as manifested variables that adjust model parameters to individual-specific values. In our case, these individual-specific values are individual measurement occasions. To our knowledge, this is the first study that demonstrates how to incorporate the definition variables in parallel growth curve models.

The developed model fills an existing gap by demonstrating how to fit a PBLSGM in the framework of individually-varying time points (ITPs) to estimate the knots, knot variances and knot-knot association. In this current work, we have three major goals. First, we aim to capture characteristics of parallel individual trajectories with linear-linear piecewise functional form and analyze the associations between multiple developmental processes. More importantly, we desire to make statistical inferences and interpret estimates in the original parameter setting of the proposed model. Second, with the definition variable approach, we fit the model in the framework of individual measurement occasions due to its omnipresence in longitudinal studies. Third, we provide a set of recommendations for real-world practice by demonstrating how to apply the proposed model to a real-world data set. 

We organize the remainder of this article as follows. In the method section, we start from a bilinear spline growth model for univariate repeated measurements. We then extend it to a parallel growth curve model and introduce the model specification of the PBLSGM to estimate knots, knot variances and knot-knot association. Next, we extend the (inverse-) transformation functions and matrices proposed by \citet{Liu2019BLSGM} to the PBLSGM framework and demonstrate how to reparameterize growth factors in this model to make them estimable and inversely transform them to the original setting so that their estimates are interpretable. Additionally, we propose a possible reduced PBLSGM for estimating knots without considering variability as a parsimonious backup. Next, we describe the model estimation and model evaluation that is realized by the Monte Carlo simulation. Then in the section of the simulation result, we present the evaluation of the model performance in terms of non-convergence rate, the proportion of improper solutions, the performance measures, which include the relative bias, the empirical standard error (SE), the relative root-mean-squared-error (RSME) and the empirical coverage probability for a nominal $95\%$ confidence interval of each parameter. By applying the proposed PBLSGM to a data set of longitudinal reading scores, mathematics scores, and science scores from Early Childhood Longitudinal Study, Kindergarten Class of $2010-11$ (ECLS-K: $2011$), we provide a collection of feasible recommendations for empirical practice. Finally, discussions are framed regarding the model's limitations as well as future directions. 

\section{Method}\label{method}
\subsection{Bilinear Spline Growth Model with a Random Knot}\label{M:BLSGM}
In this section, we briefly describe a bilinear spline growth model (BLSGM, also referred to as a linear-linear piecewise latent growth model) with a random knot, which is used for analyzing a univariate nonlinear change pattern, say the developmental process in mathematics ability, and estimating a change-point with considering variability. As an extension of latent growth curve (LGC) models, the linear-linear piecewise latent growth model specifies a separate linear function for each of the two stages of development for each individual, shown in Figure \ref{fig:knot}. In the framework of individual measurement occasions, the measurement at the $j^{th}$ time point of the $i^{th}$ individual is 
\begin{equation}\label{eq:linear-linear}
y_{ij}=\begin{cases}
\eta_{0i}^{[y]}+\eta_{1i}^{[y]}t_{ij}+\epsilon^{[y]}_{ij} & t_{ij}\le\gamma_{i}^{[y]}\\
\eta_{0i}^{[y]}+\eta_{1i}^{[y]}\gamma_{i}^{[y]}+\eta_{2i}^{[y]}(t_{ij}-\gamma_{i}^{[y]})+\epsilon^{[y]}_{ij} & t_{ij}>\gamma_{i}^{[y]}\\
\end{cases}, 
\end{equation}
where $y_{ij}$ and $t_{ij}$ are the measurement and measurement occasion of the $i^{th}$ individual at time $j$. In Equation (\ref{eq:linear-linear}), $\eta_{0i}^{[y]}$, $\eta_{1i}^{[y]}$, $\eta_{2i}^{[y]}$ and $\gamma_{i}^{[y]}$ are individual-level intercept, first slope, second slope and knot which all together determine the change pattern of the growth curve of $\boldsymbol{y}_{i}$. Accordingly, they are usually called `growth factors' in the LGC literature.

Note that Equation (\ref{eq:linear-linear}) does not fit into the LGC framework for two reasons. First of all, the measurement does not have a unified expression pre- and post-knot. More importantly, Equation (\ref{eq:linear-linear}), which specifies a nonlinear relationship between the outcome $y_{ij}$ and the growth factor $\gamma_{i}^{[y]}$, cannot be estimated in the SEM framework directly \citep{Grimm2016growth3}. The repeated measures in Equation (\ref{eq:linear-linear}) can be reparameterized \citep{Tishler1981nonlinear, Seber2003nonlinear, Grimm2016growth, Liu2019knot, Liu2019BLSGM} to have a unified expression of repeated measurements before and after the knot (see Appendix \ref{Supp:1A} for details of reparameterization). We then express the repeated measures as a linear combination of all four growth factors using the Taylor series expansion \citep{Browne1991Taylor, Grimm2016growth, Liu2019knot, Liu2019BLSGM} (see Appendix \ref{Supp:1B} for details of Taylor series expansion). That is, the model specified in Equation (\ref{eq:linear-linear}) can be written as a standard LGC model with reparameterized growth factors
\begin{equation}\label{eq:BLSGM}
\boldsymbol{y}_{i}=\boldsymbol{\Lambda}_{i}^{'[y]}\times\boldsymbol{\eta}^{'[y]}_{i}+\boldsymbol{\epsilon}^{[y]}_{i},
\end{equation}
where $\boldsymbol{y}_{i}$ is a $J\times 1$ vector of the repeated measures for the $i^{th}$ individual (in which $J$ is the number of measurements), $\boldsymbol{\eta}^{'[y]}_{i}$ is a $4\times1$ vector of reparameterized growth factors and $\boldsymbol{\Lambda}_{i}^{'[y]}$, which is a function of time points and the knot, is a $J\times4$ matrix of factor loadings. Additionally, $\boldsymbol{\epsilon}^{[y]}_{i}$ is a $J\times 1$ vector of residuals of the $i^{th}$ person. For the $i^{th}$ individual, we express the reparameterized growth factors (the measurement at the knot, the mean of two slopes, the half difference of two slopes and the deviation from the knot mean) and their factor loadings as 
\begin{equation}\label{eq:refactor_r}
\boldsymbol{\eta}_{i}^{'[y]} = \begin{pmatrix}
\eta_{0i}^{'[y]} & \eta_{1i}^{'[y]} & \eta_{2i}^{'[y]} & \delta^{[y]}_{i}
\end{pmatrix}^{T}
= \begin{pmatrix}
\eta^{[y]}_{0i}+\gamma^{[y]}_{i}\eta^{[y]}_{1i} & \frac{\eta^{[y]}_{1i}+\eta^{[y]}_{2i}}{2} & \frac{\eta^{[y]}_{2i}-\eta^{[y]}_{1i}}{2} & \gamma^{[y]}_{i}-\mu^{[y]}_{\gamma}
\end{pmatrix}^{T}
\end{equation}
and
\begin{equation}\label{eq:reloading_r}
\begin{aligned}
&\boldsymbol{\Lambda}_{i}^{'[y]} = \begin{pmatrix}
1 & t_{ij}-\mu^{[y]}_{\gamma} & |t_{ij}-\mu^{[y]}_{\gamma}| & -\mu^{'[y]}_{\eta_{2}}-\frac{\mu^{'[y]}_{\eta_{2}}(t_{ij}-\mu^{[y]}_{\gamma})}{|t_{ij}-\mu^{[y]}_{\gamma}|}
\end{pmatrix}
&(j=1,\cdots, J),
\end{aligned}
\end{equation}
respectively, where $\mu^{[y]}_{\gamma}$ is the knot mean and $\delta^{[y]}_{i}$ is the deviation from the knot mean of the $i^{th}$ individual. 

\subsection{Model Specification of Parallel Bilinear Spline Growth Models with Random Knots}\label{M:specify}
In this section, we extend the univariate BLSGM to its parallel version, which allows for analyzing joint development, say trajectories of development in reading and mathematics ability. The model specification of the PBLSGM with unknown random knots in the framework of individual measurement occasions can be realized by extending the model in Equation (\ref{eq:BLSGM}). Suppose we have bivariate growth curves of repeated outcomes $\boldsymbol{y}_{i}$ and $\boldsymbol{z}_{i}$ for each individual, the PBLSGM is given
\begin{equation}\label{eq:specify_r1}
\begin{pmatrix}
\boldsymbol{y}_{i} \\ \boldsymbol{z}_{i}
\end{pmatrix}=
\begin{pmatrix}
\boldsymbol{\Lambda}_{i}^{'[y]} & \boldsymbol{0} \\ \boldsymbol{0} & \boldsymbol{\Lambda}_{i}^{'[z]}
\end{pmatrix}\times
\begin{pmatrix}
\boldsymbol{\eta}^{'[y]}_{i} \\ \boldsymbol{\eta}^{'[z]}_{i}
\end{pmatrix}+
\begin{pmatrix}
\boldsymbol{\epsilon}^{[y]}_{i} \\ \boldsymbol{\epsilon}^{[z]}_{i}
\end{pmatrix},
\end{equation}
where $\boldsymbol{z}_{i}$ is also a $J\times 1$ vector of the repeated measures for the $i^{th}$ individual, $\boldsymbol{\eta}^{'[z]}_{i}$, $\boldsymbol{\Lambda}_{i}^{'[z]}$ and $\boldsymbol{\epsilon}^{[z]}_{i}$ are its reparameterized growth factors (a $4\times1$ vector), the corresponding factor loadings (a $J\times4$ matrix), and the residuals of the $i^{th}$ person (a $J\times 1$ vector), respectively. Same to the reparameterization of $\boldsymbol{y}_{i}$, the reparameterized growth factors of $\boldsymbol{z}_{i}$ are also the measurement of the knot, the mean of two slopes, the half difference of two slopes and the deviation from the knot mean. The outcome-specific reparameterized growth factors $\boldsymbol{\eta}^{'[u]}_{i}$ ($u=y,\ z$) can be further written as deviations from the corresponding outcome-specific means
\begin{equation}\label{eq:specify_r2}
\begin{pmatrix}
\boldsymbol{\eta}^{'[y]}_{i} \\ \boldsymbol{\eta}^{'[z]}_{i}
\end{pmatrix}=
\begin{pmatrix}
\boldsymbol{\mu}^{'[y]}_{\boldsymbol{\eta}} \\ \boldsymbol{\mu}^{'[z]}_{\boldsymbol{\eta}}
\end{pmatrix}+
\begin{pmatrix}
\boldsymbol{\zeta}^{'[y]}_{i} \\
\boldsymbol{\zeta}^{'[z]}_{i}
\end{pmatrix},
\end{equation}
where $\boldsymbol{\mu}^{'[u]}_{\boldsymbol{\eta}}$ is a $4\times 1$ vector of outcome-specific reparameterized growth factor means and $\boldsymbol{\zeta}^{'[u]}_{i}$ is a $4\times 1$ vector of deviations of the $i^{th}$ subject from the growth factor means. It is noted that $\begin{pmatrix} \boldsymbol{\zeta}^{'[y]}_{i} & \boldsymbol{\zeta}^{'[z]}_{i}\end{pmatrix}^{T}$ follows a multivariate normal distribution
\begin{equation}\label{eq:specify_r3}
\begin{pmatrix} 
\boldsymbol{\zeta}^{'[y]}_{i} \\ \boldsymbol{\zeta}^{'[z]}_{i}
\end{pmatrix}\sim \text{MVN}\bigg(\boldsymbol{0}, 
\begin{pmatrix}
\boldsymbol{\Psi}_{\boldsymbol{\eta}}^{'[y]} & \boldsymbol{\Psi}_{\boldsymbol{\eta}}^{'[yz]} \\
& \boldsymbol{\Psi}_{\boldsymbol{\eta}}^{'[z]}
\end{pmatrix}\bigg),
\end{equation}
where $\boldsymbol{\Psi}_{\boldsymbol{\eta}}^{'[u]}$ is a $4\times 4$ variance-covariance matrix of outcome-specific reparameterized growth factors and $\boldsymbol{\Psi}_{\boldsymbol{\eta}}^{'[yz]}$, which is also a $4\times4$ matrix that indicates the covariances between reparameterized growth factors of repeated outcomes $\boldsymbol{y}_{i}$ and $\boldsymbol{z}_{i}$. To simplify the model, we assume that individual outcome-specific residuals, $\boldsymbol{\epsilon}^{[u]}_{i}$ in Equation (\ref{eq:specify_r1}), are identical and independent normal distributions over time and the residual covariances are homogeneous over time, that is,
\begin{equation}\nonumber
\begin{pmatrix} 
\boldsymbol{\epsilon}^{[y]}_{i} \\ \boldsymbol{\epsilon}^{[z]}_{i}
\end{pmatrix}\sim \text{MVN}\bigg(\boldsymbol{0}, 
\begin{pmatrix}
\theta^{[y]}_{\epsilon}\boldsymbol{I} & \theta^{[yz]}_{\epsilon}\boldsymbol{I} \\
& \theta^{[z]}_{\epsilon}\boldsymbol{I}
\end{pmatrix}\bigg),
\end{equation}
where $\boldsymbol{I}$ is a $J\times J$ identity matrix.

\subsection{Transformation between Two Parameter-spaces}\label{method:matrix}
A set of reasonable initial values usually expedites the computational process and increases the likelihood of convergence of a complicated model in the SEM framework. Empirical researchers usually employ descriptive statistics and visualization techniques to choose proper initial values for the parameters. Nevertheless, for the parameters in the reparameterized frame, this decision is not straightforward. To help select initial values, \citet{Liu2019BLSGM} proposed a transformation function and matrix to re-express the mean vector and variance-covariance matrix of growth factors in the original setting as those in the reparameterized setting for the BLSGM with an unknown random knot. They also developed an inverse-transformation function and matrix to transform the mean vector and variance-covariance matrix of reparameterized growth factors to the original setting so that the estimates are interpretable. The derivation of (inverse-)transformation matrix is realized by employing the \textit{Multivariate Delta Method}. By extending that work, we demonstrate how to implement these (inverse-) transformation matrices in the PBLSGM. 

As shown in Equation (\ref{eq:refactor_r}), the relationship between the growth factors in the original setting and those in the reparameterized framework is individual-level since each individual has a `personal' set of growth factors. By utilizing the \textit{Multivariate Delta Method}, the transformation is simplified to be the population level. \citet{Liu2019BLSGM} demonstrated that the BLSGM with a random knot generally produce unbiased point estimates with small standard errors and confidence intervals with satisfactory coverage probabilities for the parameters that are directly related to the underlying change patterns. Therefore, in this current work, we still employ the population-level (inverse-) transformation for each repeated outcome. We provide detailed derivation for the transformation and inverse-transformation between the outcome-specific growth factors' mean vector and variance-covariance structure in the original setting and those in the reparameterized frame in Appendix \ref{Supp:1C}. 

When fitting a PBLSGM using the \textit{R} package \textit{OpenMx}, which allows for matrix algebra on those estimates from the model by the function \textit{mxAlgebra()} \citep{User2018OpenMx}, we only need to provide the inverse-transformation function and matrix between two parameter-spaces rather than derive the final expression of each parameter in the original setting from those in the reparameterized frame. Specifically, we need to specify the inverse-transformation function and matrix, the estimated mean vector and variance-covariance structure of reparameterized growth factors, and the algebraic expression (i.e., matrix multiplication) between them in the function \textit{mxAlgebra()}; then \textit{OpenMx} is capable of computing the point estimates along with their standard errors of the growth factor parameters in the original setting automatically. Other SEM software such as \textit{Mplus} can also compute new parameters to be derived from those estimated automatically by specifying their relationship in the \textit{NEW} command. However, \textit{Mplus} does not allow for matrix algebra; we then need to derive the expression for each cell of the mean vector and the variance-covariance matrix of the original growth factors. All these expressions are also in Appendix \ref{Supp:1C}. We provide \textit{OpenMx} code and \textit{Mplus} 8 syntax in the online appendix (\url{https://github.com/Veronica0206/Extension_projects}) to demonstrate how to transform the estimated mean vector and variance-covariance structure of reparameterized growth factors to those in the original setting. 

\subsection{Model Estimation}\label{method:est}
For the $i^{th}$ individual, we then write the expected mean vector and variance-covariance matrix of the bivariate repeated outcomes of the PBLSGM specified in Equation (\ref{eq:specify_r1}) as
\begin{equation}\label{eq:mean_r}
\boldsymbol{\mu}_{i}=\begin{pmatrix}
\boldsymbol{\mu}^{[y]}_{i} \\ \boldsymbol{\mu}^{[z]}_{i}
\end{pmatrix}=\begin{pmatrix}
\boldsymbol{\Lambda}_{i}^{'[y]} & \boldsymbol{0} \\ \boldsymbol{0} & \boldsymbol{\Lambda}_{i}^{'[z]}
\end{pmatrix}\times\begin{pmatrix}
\boldsymbol{\mu}^{'[y]}_{\boldsymbol{\eta}} \\ \boldsymbol{\mu}^{'[z]}_{\boldsymbol{\eta}}
\end{pmatrix}
\end{equation}
and
\begin{equation}\label{eq:var_r}
\begin{aligned}
\boldsymbol{\Sigma}_{i}&=\begin{pmatrix}
\boldsymbol{\Sigma}^{[y]}_{i} & \boldsymbol{\Sigma}^{[yz]}_{i} \\
& \boldsymbol{\Sigma}^{[z]}_{i}
\end{pmatrix}\\
&=\begin{pmatrix}
\boldsymbol{\Lambda}_{i}^{'[y]} & \boldsymbol{0} \\ \boldsymbol{0} & \boldsymbol{\Lambda}_{i}^{'[z]}
\end{pmatrix}\times\begin{pmatrix}
\boldsymbol{\Psi}^{'[y]}_{\boldsymbol{\eta}} & \boldsymbol{\Psi}^{'[yz]}_{\boldsymbol{\eta}} \\
& \boldsymbol{\Psi}^{'[z]}_{\boldsymbol{\eta}}
\end{pmatrix}\times\begin{pmatrix}
\boldsymbol{\Lambda}_{i}^{'[y]} & \boldsymbol{0} \\ \boldsymbol{0} & \boldsymbol{\Lambda}_{i}^{'[z]}
\end{pmatrix}^{T}+\begin{pmatrix}
\theta^{[y]}_{\epsilon}\boldsymbol{I} & \theta^{[yz]}_{\epsilon}\boldsymbol{I} \\
& \theta^{[z]}_{\epsilon}\boldsymbol{I}
\end{pmatrix}.
\end{aligned}
\end{equation}

The parameters in the PBLSGM specified in Equations (\ref{eq:specify_r2}) and (\ref{eq:specify_r3}) include the mean vector and variance-covariance matrix of outcome-specific reparameterized growth factors, the reparameterized growth factor covariances, the outcome-specific residual variance, and the residual covariance. Then we can calculate the parameters in the original setting with the inverse-transformation function and matrix proposed in Section \ref{method:matrix}. $\boldsymbol{\Theta}_{1}$ and $\boldsymbol{\Theta}_{1}^{'}$ defined as 
\begin{equation}\label{eq:theta_r}
\begin{aligned}
\boldsymbol{\Theta}_{1}=&\{\boldsymbol{\mu}^{[u]}_{\boldsymbol{\eta}}, \boldsymbol{\Psi}^{[u]}_{\boldsymbol{\eta}},  \boldsymbol{\Psi}^{[yz]}_{\boldsymbol{\eta}},  \theta^{[u]}_{\epsilon}, \theta^{[yz]}_{\epsilon}\}\\
=&\{\mu^{[u]}_{\eta_{0}}, \mu^{[u]}_{\eta_{1}}, \mu^{[u]}_{\eta_{2}}, \mu^{[u]}_{\gamma}, \psi^{[u]}_{00}, \psi^{[u]}_{01}, \psi^{[u]}_{02}, \psi^{[u]}_{0\gamma}, \psi^{[u]}_{11}, \psi^{[u]}_{12}, \psi^{[u]}_{1\gamma}, \psi^{[u]}_{22}, \psi^{[u]}_{2\gamma}, \psi^{[u]}_{\gamma\gamma}, \\
&\psi^{[yz]}_{00}, \psi^{[yz]}_{01}, \psi^{[yz]}_{02}, \psi^{[yz]}_{0\gamma}, \psi^{[yz]}_{10}, \psi^{[yz]}_{11}, \psi^{[yz]}_{12}, \psi^{[yz]}_{1\gamma},\psi^{[yz]}_{20}, \psi^{[yz]}_{21}, \psi^{[yz]}_{22}, \psi^{[yz]}_{2\gamma}, \ \ \ \ \ \ \ \ \ \ \ \\
& \psi^{[yz]}_{\gamma 0}, \psi^{[yz]}_{\gamma 1}, \psi^{[yz]}_{\gamma 2}, \psi^{[yz]}_{\gamma\gamma}, \theta^{[u]}_{\epsilon}, \theta^{[yz]}_{\epsilon}\}\\
&u=y,\ z
\end{aligned}
\end{equation}
and
\begin{equation}\label{eq:retheta_r}
\begin{aligned}
\boldsymbol{\Theta}^{'}_{1}=&\{\boldsymbol{\mu}^{'[u]}_{\boldsymbol{\eta}}, \boldsymbol{\Psi}^{'[u]}_{\boldsymbol{\eta}}, \boldsymbol{\Psi}^{'[yz]}_{\boldsymbol{\eta}},  \theta^{[u]}_{\epsilon}, \theta^{[yz]}_{\epsilon}\}\\
=&\{\mu^{'[u]}_{\eta_{0}}, \mu^{'[u]}_{\eta_{1}}, \mu^{'[u]}_{\eta_{2}}, \mu^{'[u]}_{\gamma}, \psi^{'[u]}_{00}, \psi^{'[u]}_{01}, \psi^{'[u]}_{02}, \psi^{'[u]}_{0\gamma}, \psi^{'[u]}_{11}, \psi^{'[u]}_{12}, \psi^{'[u]}_{1\gamma}, \psi^{'[u]}_{22}, \psi^{'[u]}_{2\gamma}, \psi^{'[u]}_{\gamma\gamma}, \\
&\psi^{'[yz]}_{00}, \psi^{'[yz]}_{01}, \psi^{'[yz]}_{02}, \psi^{'[yz]}_{0\gamma}, \psi^{'[yz]}_{10}, \psi^{'[yz]}_{11}, \psi^{'[yz]}_{12}, \psi^{'[yz]}_{1\gamma},\psi^{'[yz]}_{20}, \psi^{'[yz]}_{21}, \psi^{'[yz]}_{22}, \psi^{'[yz]}_{2\gamma}, \\
& \psi^{'[yz]}_{\gamma 0}, \psi^{'[yz]}_{\gamma 1}, \psi^{'[yz]}_{\gamma 2}, \psi^{'[yz]}_{\gamma\gamma}, \theta^{[u]}_{\epsilon}, \theta^{[yz]}_{\epsilon}\} \\
&u=y,\ z,
\end{aligned}
\end{equation}
which are the parameters in the original setting and those in the reparameterized frame. 

$\boldsymbol{\Theta}_{1}^{'}$ is estimated using full information maximum likelihood (FIML) to account for the potential heterogeneity of individual contributions to the likelihood function. We then express the individual-level and sample-level log-likelihood function as
\begin{equation}\label{eq:loglik1}
\log lik_{i}(\boldsymbol{\Theta}_{1}^{'}|\boldsymbol{y}_{i}, \boldsymbol{z}_{i})=C-\frac{1}{2}\ln\begin{vmatrix}
\boldsymbol{\Sigma}^{[y]}_{i} & \boldsymbol{\Sigma}^{[yz]}_{i} \\
& \boldsymbol{\Sigma}^{[z]}_{i}
\end{vmatrix}-\frac{1}{2}\bigg(\begin{pmatrix}
\boldsymbol{y}_{i} - \boldsymbol{\mu}^{[y]}_{i} \\ \boldsymbol{z}_{i} - \boldsymbol{\mu}^{[z]}_{i}
\end{pmatrix}^{T}\begin{pmatrix}
\boldsymbol{\Sigma}^{[y]}_{i} & \boldsymbol{\Sigma}^{[yz]}_{i} \\
& \boldsymbol{\Sigma}^{[z]}_{i}
\end{pmatrix}^{-1}\begin{pmatrix}
\boldsymbol{y}_{i} - \boldsymbol{\mu}^{[y]}_{i} \\ \boldsymbol{z}_{i} - \boldsymbol{\mu}^{[z]}_{i}
\end{pmatrix}\bigg)
\end{equation}
and
\begin{equation}\label{eq:loglik2}
\log lik(\boldsymbol{\Theta}_{1}^{'})=\sum_{i=1}^{n}\log lik_{i}(\boldsymbol{\Theta}_{1}^{'}|\boldsymbol{y}_{i}, \boldsymbol{z}_{i}),
\end{equation}
respectively,  where $n$ is the number of individuals, and $C$ is a constant. In this work, the proposed PBLSGM is built using the R package \textit{OpenMx} with CSOLNP optimizer \citep{Pritikin2015OpenMx, OpenMx2016package, User2018OpenMx, Hunter2018OpenMx}. One advantage of \textit{OpenMx} lies in that it allows for matrix calculation so that we can carry out the inverse-transformation function and matrix proposed in Section \ref{method:matrix} efficiently. In the online appendix, we provide the \textit{OpenMx} code for the proposed PBLSGM and a demonstration. \textit{Mplus} 8 code is also provided for the model in the online appendix for researchers who are interested in using \textit{Mplus}.

\subsection{Reduced Model}\label{method:reduced}
A reduced model can be created with the assumption that the outcome-specific change-point is roughly similar for each individual. We then fix the between-individual differences in each knot to $0$ and build a PBLSGM for estimating unknown knots without variability as a reduced form of the model in Equation (\ref{eq:specify_r1}). It is given by
\begin{equation}\label{eq:specify_f1}
\begin{pmatrix}
\boldsymbol{y}_{i} \\ \boldsymbol{z}_{i}
\end{pmatrix}=
\begin{pmatrix}
\boldsymbol{\Lambda}_{i}^{'[y]} & \boldsymbol{0} \\ \boldsymbol{0} & \boldsymbol{\Lambda}_{i}^{'[z]}
\end{pmatrix}\times
\begin{pmatrix}
\boldsymbol{\eta}^{'[y]}_{i} \\ \boldsymbol{\eta}^{'[z]}_{i}
\end{pmatrix}+
\begin{pmatrix}
\boldsymbol{\epsilon}^{[y]}_{i} \\ \boldsymbol{\epsilon}^{[z]}_{i}
\end{pmatrix},
\end{equation}
where $\boldsymbol{\eta}^{'[u]}_{i}$ is a $3\times1$ vector of outcome-specific reparameterized growth factors (the measurement at the knot, the mean of two slopes, and the half difference between two slopes), and $\boldsymbol{\Lambda}_{i}^{'[u]}$, a function of time points and the outcome-specific fixed knot $\gamma^{[u]}$, is a $J\times 3$ matrix of individual-specific factor loadings. The outcome-specific reparameterized growth factors $\boldsymbol{\eta}^{'[u]}_{i}$ and corresponding factor loadings $\boldsymbol{\Lambda}_{i}^{'[u]}$ can be further expressed as 
\begin{equation}\label{eq:refactor_f}
\boldsymbol{\eta}_{i}^{'[u]} = \left(\begin{array}{rrr}
\eta_{0i}^{'[u]} & \eta_{1i}^{'[u]} & \eta_{2i}^{'[u]} 
\end{array}\right)^{T}
= \left(\begin{array}{rrr}
\eta^{[u]}_{0i}+\gamma^{[u]}\eta^{[u]}_{1i} & \frac{\eta^{[u]}_{1i}+\eta^{[u]}_{2i}}{2} & \frac{\eta^{[u]}_{2i}-\eta^{[u]}_{1i}}{2} 
\end{array}\right)^{T}
\end{equation}
and
\begin{equation}\label{eq:reloading_f}
\begin{aligned}
&\boldsymbol{\Lambda}_{i}^{'[u]} = \left(\begin{array}{rrr}
1 & t_{ij}-\gamma^{[u]} & |t_{ij}-\gamma^{[u]}| 
\end{array}\right)
&(j=1,\cdots, J).
\end{aligned}
\end{equation}

The mean vector ($\boldsymbol{\mu}^{'[u]}_{\boldsymbol{\eta}}$) and variance-covariance matrix ($\boldsymbol{\Psi}^{'[u]}_{\boldsymbol{\eta}}$) of the outcome-specific growth factors and between-construct growth factor covariance matrix  ($\boldsymbol{\Psi}^{'[yz]}_{\boldsymbol{\eta}}$) also reduce to be a $3\times1$ vector, $3\times3$ matrix and $3\times3$ matrix, respectively. We also need to reduce the (inverse-) transformation function and matrix accordingly. Specifically, we only need the first three entries of the (inverse-) transformation function as well as the first three columns and the first three rows of the (inverse-) transformation block matrices, since only the intercept and two slopes for each repeated outcome need to be reparameterized. A demonstration of implementing the (inverse-) transformation in the reduced model is also provided in the online appendix. 

For the $i^{th}$ individual, with the reparameterized growth factors and their factor loadings as defined in Equations (\ref{eq:refactor_f}) and (\ref{eq:reloading_f}), the expected mean vector and the variance-covariance matrix of the bivariate repeated measurements of the reduced PBLSGM are given by
\begin{equation}\label{eq:mean_f}
\boldsymbol{\mu}_{i}=\begin{pmatrix}
\boldsymbol{\mu}^{[y]}_{i} \\ \boldsymbol{\mu}^{[z]}_{i}
\end{pmatrix}=\begin{pmatrix}
\boldsymbol{\Lambda}_{i}^{'[y]} & \boldsymbol{0} \\ \boldsymbol{0} & \boldsymbol{\Lambda}_{i}^{'[z]}
\end{pmatrix}\times\begin{pmatrix}
\boldsymbol{\mu}^{'[y]}_{\boldsymbol{\eta}} \\ \boldsymbol{\mu}^{'[z]}_{\boldsymbol{\eta}}
\end{pmatrix}
\end{equation}
and
\begin{equation}\label{eq:var_f}
\begin{aligned}
\boldsymbol{\Sigma}_{i}&=\begin{pmatrix}
\boldsymbol{\Sigma}^{[y]}_{i} & \boldsymbol{\Sigma}^{[yz]}_{i} \\
& \boldsymbol{\Sigma}^{[z]}_{i}
\end{pmatrix}\\
&=\begin{pmatrix}
\boldsymbol{\Lambda}_{i}^{'[y]} & \boldsymbol{0} \\ \boldsymbol{0} & \boldsymbol{\Lambda}_{i}^{'[z]}
\end{pmatrix}\times\begin{pmatrix}
\boldsymbol{\Psi}^{'[y]}_{\boldsymbol{\eta}} & \boldsymbol{\Psi}^{'[yz]}_{\boldsymbol{\eta}} \\
& \boldsymbol{\Psi}^{'[z]}_{\boldsymbol{\eta}}
\end{pmatrix}\times\begin{pmatrix}
\boldsymbol{\Lambda}_{i}^{'[y]} & \boldsymbol{0} \\ \boldsymbol{0} & \boldsymbol{\Lambda}_{i}^{'[z]}
\end{pmatrix}^{T}+\begin{pmatrix}
\theta^{[y]}_{\epsilon}\boldsymbol{I} & \theta^{[yz]}_{\epsilon}\boldsymbol{I} \\
& \theta^{[z]}_{\epsilon}\boldsymbol{I}
\end{pmatrix},
\end{aligned}
\end{equation}
respectively. For the reduced model, $\boldsymbol{\Theta}_{2}$ and $\boldsymbol{\Theta}_{2}^{'}$ are defined as
\begin{equation}\label{eq:theta_f}
\begin{aligned}
\boldsymbol{\Theta}_{2}=&\{\boldsymbol{\mu}^{[u]}_{\boldsymbol{\eta}}, \boldsymbol{\Psi}^{[u]}_{\boldsymbol{\eta}}, \boldsymbol{\Psi}^{[yz]}_{\boldsymbol{\eta}}, \theta^{[u]}_{\epsilon}, \theta^{[yz]}_{\epsilon}\}\\
=&\{\mu^{[u]}_{\eta_{0}}, \mu^{[u]}_{\eta_{1}}, \mu^{[u]}_{\eta_{2}}, \gamma^{[u]}, \psi^{[u]}_{00}, \psi^{[u]}_{01}, \psi^{[u]}_{02}, \psi^{[u]}_{11}, \psi^{[u]}_{12}, \psi^{[u]}_{22},\\
&\psi^{[yz]}_{00}, \psi^{[yz]}_{01}, \psi^{[yz]}_{02}, \psi^{[yz]}_{10}, \psi^{[yz]}_{11}, \psi^{[yz]}_{12}, \psi^{[yz]}_{20}, \psi^{[yz]}_{21}, \psi^{[yz]}_{22}, \theta^{[u]}_{\epsilon}, \theta^{[yz]}_{\epsilon}\}\ \ \ \ \ \ \ \\
&u=y,\ z
\end{aligned}
\end{equation}
and
\begin{equation}\label{eq:retheta_f}
\begin{aligned}
\boldsymbol{\Theta}^{'}_{2}=&\{\boldsymbol{\mu}^{'[u]}_{\boldsymbol{\eta}}, \boldsymbol{\Psi}^{'[u]}_{\boldsymbol{\eta}}, \boldsymbol{\Psi}^{'[yz]}_{\boldsymbol{\eta}}, \theta^{[u]}_{\epsilon}, \theta^{[yz]}_{\epsilon}\}\\
=&\{\mu^{'[u]}_{\eta_{0}}, \mu^{'[u]}_{\eta_{1}}, \mu^{'[u]}_{\eta_{2}}, \gamma^{[u]}, \psi^{'[u]}_{00}, \psi^{'[u]}_{01}, \psi^{'[u]}_{02}, \psi^{'[u]}_{11}, \psi^{'[u]}_{12}, \psi^{'[u]}_{22}, \\
&\psi^{'[yz]}_{00}, \psi^{'[yz]}_{01}, \psi^{'[yz]}_{02}, \psi^{'[yz]}_{10}, \psi^{'[yz]}_{11}, \psi^{'[yz]}_{12}, \psi^{'[yz]}_{20}, \psi^{'[yz]}_{21}, \psi^{'[yz]}_{22}, \theta^{[u]}_{\epsilon}, \theta^{[yz]}_{\epsilon}\}\\
&u=y,\ z
\end{aligned}
\end{equation}
for the parameters in the original setting and those in the reparameterized frame, respectively. We then update $\boldsymbol{\mu}_{i}$ and $\boldsymbol{\Sigma}_{i}$ as those defined in Equations (\ref{eq:mean_f}) and (\ref{eq:var_f}) and obtain the individual-level and sample-level log-likelihood function to estimate $\boldsymbol{\Theta}_{2}^{'}$. The reduced PBLSGM is also constructed using the R package \textit{OpenMx} with CSOLNP optimizer and the parameters are estimated with the FIML technique. \textit{R} code and \textit{Mplus} 8 syntax of the reduced PBLSGM are also provided in the online Appendix. 

\section{Model Evaluation}\label{Evaluation}
By Monte Carlo simulation studies, we evaluated the proposed PBLSGM with two goals. The first goal was to examine the performance measures of the proposed PBLSGM, including the relative bias, the empirical standard error (SE), the relative root-mean-square error (RMSE), and the empirical coverage probability for a nominal $95\%$ confidence interval of each parameter. Table \ref{tbl:metric} lists the definitions and estimates of these four performance measures. Specifically, the relative bias quantifies whether the model targets $\theta$ (the true value of the parameter of interest) on average. The empirical SE is a metric of the precision of the estimator of $\theta$. The relative RMSE is a natural way to integrate the bias and the precision measure into one metric. The coverage probability of confidence intervals quantifies how well the interval estimate covers the parameter's population value. The second goal was to see how the reduced model performed as a parsimonious backup of the full PBLSGM. 

\tablehere{1}

Following \citet{Morris2019simulation}, we decided the number of repetitions $S=1,000$ by an empirical method. The performance metric of the highest importance in the simulation study was the (relative) bias. According to a pilot simulation run, standard errors of all parameters except the intercept variances and covariance (i.e., $\psi_{00}^{[u]}$ and $\psi_{00}^{[yz]}$) were less than $0.15$. To keep the Monte Carlo standard error\footnote{$\text{Monte Carlo SE(Bias)}=\sqrt{Var(\hat{\theta})/S}$ \citep{Morris2019simulation}.} of bias lower than $0.005$, we needed at least $900$ repetitions. We then decided to proceed with $S=1,000$ to be relatively conservative. 

\subsection{Design of Simulation Study}\label{Evaluation:design}
As mentioned earlier, the parameters of the most interest in the proposed model are the knots, knot variances, and knot-knot covariance. The conditions hypothesized to influence the estimation of these knot parameters, along with other model parameters, included sample size, the number of repeated measurements, the knot locations, the correlation of between-construct growth factors, shapes of trajectories, and measurement precision. 

Table \ref{tbl:simu_design} lists all conditions we considered in the simulation design. For a model to analyze longitudinal data, the most interesting factor is the number of repeated measures. Generally, the model should perform better with an increasing number of repeated measurements, which we wanted to examine by the simulation study. For the linear-linear functional form of trajectories, another important factor is the knot location. Intuitively, the model should perform the best when the knot is in the middle of study duration; we were interested in testing this hypothesis. We also realized that the knot locations of the two constructs (similar vs. different) may affect the proposed model, which we desired to explore through the simulation study. Accordingly, we chose two different levels of the number of measurements: $6$ and $10$. We selected $6$ as the minimum number of repeated outcomes to make the proposed model fully identified\footnote{It has been proved that the bilinear latent growth model can be identified with at least five waves with a specified knot at the midway of study duration \citep{Bollen2005LCM}, though the information of model identification of BLSGM with an unknown knot lacks.}. For the conditions with $6$ repeated measures, we set both knots at halfway of study duration ($\mu_{\gamma}^{[u]}=2.5$). The other conditon, $10$ measurement occasions, was considered for two reasons. First, we wanted to evaluate whether an increasing number of repeated measures would improve model performance. More importantly, $10$ measurements allowed us to place knots at different locations, say the knot of repeated outcome $\boldsymbol{y}_{i}$ and $\boldsymbol{z}_{i}$ are left-shifted ($\mu_{\gamma}^{[y]}=3.5$) and right-shifted ($\mu_{\gamma}^{[z]}=5.5$) from the middle point of study duration, in addition to placing both in the middle ($\mu_{\gamma}^{[u]}=4.5$) so that we could investigate whether such change of knot locations would affect the proposed model. Around each wave, we allowed a time-window with width $(-0.25, +0.25)$, which is a `medium' deviation as in \citet{Coulombe2015ignoring}, for individual measurement occasions. 

\tablehere{2}

As the proposed model is for joint development, how the correlation between two trajectories affects the model performance is worth exploring. We considered three levels of the between-construct growth factor correlation, $\pm{0.3}$ and $0$, for this exploration. The condition with correlation $\pm{0.3}$ was for the positive and negative moderate association, through which we wanted to see whether the sign of the correlation would affect model performance. With zero correlation conditions where the proposed model should not be applied, it was of interest to investigate how this model misspecification affects the performance of the proposed model. Additionally, we examined several common change patterns, as shown in Table \ref{tbl:simu_design} (Scenario 1, 2 and 3). For each scenario, we changed knot locations and one growth factor while fixed the other two growth factors to investigate how the trajectory shape impacts the model. We were also interested in examining the influence of the measurement precision and sample size on the model performance. Accordingly, we considered $\theta^{[u]}_{\epsilon}=1$ or $\theta^{[u]}_{\epsilon}=2$ as two levels of homogeneous outcome-specific residual variances and set the residual correlation as $0.3$. We also assessed the model at two levels of sample size, $n=200$ and $n=500$.

As an extension of an existing BLSGM proposed by \citet{Liu2019BLSGM}, the primary focus of this study is to investigate how the performance of the PBLSGM affected by the correlation between the growth patterns of two repeated outcomes. Accordingly, we did not investigate some conditions, such as the knot variance and the standardized difference between two slopes, which have demonstrated clear patterns with the model performance of BLSGM in \citet{Liu2019BLSGM}. Instead, we set the knot standard deviation as $0.3$ to be a moderate level of individual difference in each knot, and we adjusted the standardized difference between two slopes to satisfy other conditions that are of more interest in this study. We also fixed the variance-covariance matrix of within-construct growth factors since this matrix usually changes with the measurement scale and the time scale. Besides, the outcome-specific growth factors were set to be positively correlated to a moderate degree $(\rho=0.3)$. Additionally, guided by several existing studies \citep{Bauer2003GMM, Kohli2011PLGC, Kohli2015PLGC2, Liu2019BLSGM}, we kept the index of dispersion $(\sigma^{2}/\mu)$ of each growth factor at a tenth scale. 

\subsection{Data Generation and Simulation Step}\label{evaluation:step}
For every condition in Table \ref{tbl:simu_design}, we carried out the following general steps for the simulation study of the proposed PBLSGM:
\begin{enumerate}
    \item Generated growth factors for two repeated outcomes simultaneously with the prespecified mean vector and variance-covariance matrix shown in Table \ref{tbl:simu_design} using the R package \textit{MASS} \citep{Venables2002Statistics},
    \item Generated a scaled and equally-spaced time structure with $J$ waves $t_{j}$ and obtained individual measurement occasions: $t_{ij}\sim U(t_{j}-\Delta, t_{j}+\Delta)$ by allowing a time-window with width $(-\Delta, \Delta)$ around each wave,
    \item Calculated factor loadings for each individual of each construct from ITPs and the outcome-specific knot location,
    \item Calculated the values of the bivariate repeated measurements from corresponding growth factors, factor loadings, residual variances and covariance,
    \item Implemented both PBLSGMs on the generated data set, estimated the parameters, and constructed corresponding $95\%$ Wald CIs,
    \item Repeated the steps as mentioned above until achieving $1,000$ convergent solutions.
\end{enumerate}

\section{Results}\label{results}
\subsection{Model Convergence and Proper Solution}\label{result:Preliminary}
We first investigated the convergence\footnote{In this work, we defined the convergence as achieving \textit{OpenMx} status code $0$, which indicates a successful optimization, until up to $10$ attempts with different collections of initial values \citep{OpenMx2016package}.} rate and the proportion of improper solutions for every condition before evaluating how the proposed PBLSGM performed. The proposed PBLSGM and its reduced model converged satisfactorily (the convergence rate of the full PBLSGM achieved at least $95\%$ for all conditions while that of its reduced version was $100\%$). Out of a total of $108$ conditions, $57$ conditions reported a $100\%$ convergence rate and $33$ conditions reported a convergence rate of $99\%$ to $100\%$. The worst scenario in terms of the non-convergence rate was $49/1049$, indicating that we needed to repeat the steps described in Section \ref{evaluation:step} $1,049$ times to achieve $1,000$ replications with a convergent solution. The non-convergent solutions occurred under challenging conditions such as shorter study duration (i.e., $J=6$) and less precise measurements (i.e., $\theta^{[u]}_{\epsilon}=2$), or the conditions with zero between-construct growth factor covariances (i.e., $\rho=0$).

We also examined the pattern of improper solutions, which include negative estimated variances of growth factors and/or out-of-range (i.e., beyond $[-1, 1]$) correlations between growth factors. Table \ref{tbl:Improper} presents the occurrences of improper solutions produced by the proposed PBLSGM under conditions with $10$ repeated measures, including negative knot variances and out-of-range (i.e., out of $[-1, 1]$) knot correlations with any other growth factors from the same or the other construct. From the table, we noted that the conditions with real parallel trajectories of two repeated outcomes (i.e., the conditions with Scenario 1 of which both slopes of one construct were set as the same with the other construct) suffered improper solutions less frequently, though different first slopes (i.e., the conditions with Scenario 2) or second slopes (i.e., the conditions with Scenario 3) only affected the proportion of improper solution slightly. Additionally, the proposed PBLSGM was more likely to produce improper solutions under the conditions with the smaller sample size (i.e., $n=200$), less precise measurements (i.e., $\theta^{[u]}_{\epsilon}=2$), or shorter study duration (i.e., $J=6$). We replaced the PBLSGM with its reduced model for the model evaluation when such improper solutions emerged.

\tablehere{3}

\subsection{Performance Measures}\label{result:Primary}
In this section, we present simulation results in terms of performance measures, including the relative bias, empirical SE, relative RMSE and empirical coverage probability for a nominal $95\%$ confidence interval for each parameter. In general, the proposed PBLSGM is capable of estimating parameters unbiasedly, precisely, and exhibiting appropriate confidence interval coverage. Given the size of parameters and simulation conditions, we first provide the summary statistics (specifically, median and range) for each performance metric of each parameter of interest across conditions; we then discuss how these performance metrics were influenced by simulation conditions.

For the proposed PBLSGM and the reduced model, we present the median (range) of the relative bias and empirical SE of each parameter of interest across all conditions with $10$ repeated measurements in Tables \ref{tbl:rBias10_discription} and \ref{tbl:empSE10_discription}, respectively. For each parameter of interest, we first estimated its relative bias/empirical SE over $1,000$ replications under each condition with $10$ repeated measures in the simulation design. We then summarized these relative biases/empirical SEs across conditions as the corresponding median (range). 

\tablehere{4}

\tablehere{5}

From Tables \ref{tbl:rBias10_discription} and \ref{tbl:empSE10_discription}, we can see that the proposed PBLSGMs generally yielded unbiased point estimates along with small empirical standard errors and the full model performed better than the reduced model as the relative bias ranges of the parameters from the full PBLSGM were narrower than those from its reduced version. Specifically, for the full model, the magnitude of relative biases of the growth factor means was under $0.005$, and that of intercept variances and slope variances was under $0.072$. From Table \ref{tbl:rBias10_discription}, the proposed model may produce biased estimates for the knot variances and covariance: the median values of relative biases of $\psi_{\gamma\gamma}^{[y]}$, $\psi_{\gamma\gamma}^{[z]}$ and $\psi_{\gamma\gamma}^{[yz]}$ were $-0.1416$, $0.1689$ and $-0.2634$, respectively. 

To further investigate the relative bias pattern for each knot variance, we plotted the relative bias under each condition with $10$ repeated measures for $\psi_{\gamma\gamma}^{[y]}$ and $\psi_{\gamma\gamma}^{[z]}$ in Figures \ref{fig:rBiasY} and \ref{fig:rBiasZ}, respectively. From these figures, we observed how the conditions we set up in the simulation design affected these estimates. First, under the conditions where both slopes of the two constructs were the same (i.e., the conditions with Scenario 1), estimates were less biased, and both $\psi_{\gamma\gamma}^{[y]}$ and $\psi_{\gamma\gamma}^{[z]}$ were underestimated. However, under the conditions with Scenario 2 and Scenario 3, the estimates of $\psi_{\gamma\gamma}^{[y]}$ and $\psi_{\gamma\gamma}^{[z]}$ were downward and upward, respectively. Second, the downward estimates were mainly from the conditions with the small sample size (i.e., $n=200$) or less precise measurements (i.e., $\theta^{[u]}_{\epsilon}=2$) while the upward estimates were mainly from the conditions with the small residual variance.

\figurehere{2}

From Table \ref{tbl:empSE10_discription}, estimates from both full and reduced PBLSGM were precise: the magnitude of empirical standard errors of slope or knot parameters were under $0.15$ although those values of intercept parameters were relatively large. The relatively large intercept parameters' empirical SEs were due to the large scale of their population values (the true value of intercept means was around $100$, and that of intercept variances was $25$).

Table \ref{tbl:rRMSE10_discription} lists the median (range) of relative RMSE of each parameter for both models under the conditions with $10$ repeated measures, which combines bias and precision to examine the point estimate holistically. From the table, both models can estimate parameters accurately. The magnitude of relative RMSEs of growth factor means was under $0.05$, and for intercept and slope variances was under $0.20$. The relative RMSE magnitude of knot variances was relatively large due to their relatively biased point estimates.

\tablehere{6}

Table \ref{tbl:CP10_discription} presents the median (range) of the coverage probability (CP) of each parameter of interest for the proposed PBLSGM and its reduced version. Overall, the full model performed well regarding empirical coverage as the values of coverage probabilities of all parameters were near $0.95$ though the coverage probabilities produced by the reduced model were slightly less satisfied. We noticed that the coverage probabilities of knot variances and covariance may achieve $99\%$ that is greater than the nominal coverage probability ($95\%$). One reason for this phenomenon is that the empirical SEs of the knot variances and covariance were smaller than their estimated SEs. Focusing on the replications with proper solutions, the empirical SEs of $\psi_{\gamma\gamma}^{[y]}$, $\psi_{\gamma\gamma}^{[z]}$ and $\psi_{\gamma\gamma}^{[yz]}$ were around $0.041$, $0.041$ and $0.032$, respectively, while the mean of estimated SEs were around $0.045$, $0.045$ and $0.034$, respectively.

\tablehere{7}

To summarize, based on our simulation study, the estimates from the proposed PBLSGMs were unbiased and precise, with proper $95\%$ coverage probabilities generally. Some factors, for example, sample size and measurement precision, influenced model performance. Specifically, the larger sample size (i.e., $n=500$) and more precise measurement ($\theta^{[u]}_{\epsilon}=1$) improved model performance. The trajectory shape did not affect the biases meaningfully other than its influence on the relative bias of knot variances, as shown in Figure \ref{fig:rBias}. Additionally, longer study duration ($J=10$) also improved model performance (detailed performance measures under conditions with $6$ repeated measures are provided in Appendix \ref{Supp:2}). Other conditions, such as the magnitude or sign of the between-construct correlation, did not affect performance measures meaningfully.

\section{Application}\label{application}
This section demonstrates how to employ the proposed PBLSGM to analyze real-world data sets to approximate nonlinear parallel trajectories as well as estimate an outcome-specific knot with its variability and knot-knot association. This application section includes two examples. In the first example, we illustrate the recommended steps to construct the PBLSGM in practice. In the second example, we demonstrate how to apply the proposed model to joint development with a more complicated data structure where two repeated outcomes have varied study duration. We extracted $400$ students randomly from the Early Childhood Longitudinal Study Kindergarten Cohort: 2010-2011 (ECLS-K: 2011) with complete records of repeated reading IRT scaled scores, mathematics IRT scaled scores, science IRT scaled scores and age at each wave\footnote{The total sample size of ECLS-K: 2011 is $n=18174$. The number of entries after removing records with missing values (i.e., rows with any of NaN/-9/-8/-7/-1) is $n=2290$.}.

ECLS-K: 2011 is a nationwide longitudinal study of US children registered in about $900$ kindergarten programs starting from $2010-2011$ school year. In ECLS-K: 2011, children's reading ability and mathematics ability were assessed in nine waves: fall and spring of kindergarten ($2010-2011$), first ($2011-2012$) and second ($2012-2013$) grade, respectively, as well as spring of $3^{rd}$ ($2014$), $4^{th}$ ($2015$) and $5^{th}$ ($2016$) grade, respectively. Only about $30\%$ students were evaluated in the fall of $2011$ and $2012$ \citep{Le2011ECLS}. Students' science assessment started from the spring of kindergarten, and it was evaluated in eight waves accordingly. In the analysis, we used children's age (in months) instead of their grade-in-school to obtain individual measurement occasions. In the subsample, $54\%$ and $46\%$ of children were boys and girls. Additionally, $48\%$, $5\%$, $34\%$, $8\%$ and $5\%$ of students were White, Black, Hispanic, Asian and others. In this section, we construct PBLSGMs to analyze the joint development of reading and mathematics ability and that of mathematics and science skills. 

\subsection{Univariate Development}
Following \citet{Blozis2008MGM}, we fit a univariate latent growth curve model to analyze each process in isolation before constructing parallel growth models. Specifically, we built BLSGMs with an unknown knot (either fixed or random) and three models with common parametric trajectories: linear, quadratic and Jenss-Bayley for each developmental process (i.e., reading, mathematics and science). Figure \ref{fig:all_traj} presents the model implied curves on the smooth lines for each ability of each model. The nonlinear functional forms generally fit better than the linear function. Additionally, for reading ability and mathematics ability at the early stage of the ECLS-K: 2011 survey, the discrepancy between the model-implied trajectory and the smooth line of the raw data was small, suggesting that the BLSGM fit better than the models with a parametric functional form.

\figurehere{3}

Table \ref{tbl:info_uni} lists the estimated likelihood, information criteria including AIC and BIC, and residual variance of each model of each developmental process. For reading ability and science ability, the full BLSGM (i.e., the BLSGM with an unknown random knot) had the largest estimated likelihood, the smallest AIC and BIC, as well as the smallest residual variances, which led unequivocally to the selection of the full BLSGM as the `best' model from the statistical perspective. We then applied the proposed PBLSGM to analyze joint development.

\tablehere{8}

\subsection{Joint Development in Reading Ability and Mathematics Ability}
In this section, we analyzed the joint development of reading ability and mathematics ability. We first constructed the full parallel bilinear spline growth curve model and its reduced version as well as three parallel trajectory models with parametric functional forms: parallel linear, quadratic and Jenss-Bayley in the framework of individual measurement occasions where the time metric was students' age. Noting that an alternative time scale can also be grade-in-school, where the time metric takes on discrete values (for example, $0$ is for the kindergarten fall semester, and $0.5$ in years or $6$ in months is for the kindergarten spring semester, etc.), we conducted a sensitivity analysis where the PBLSGM was built with the grade-in-school as the time scale.

\subsubsection*{Main Analysis}
As shown in Figure \ref{fig:RM_traj}, the model implied curves of the parallel models did not change meaningfully from those of univariate growth models shown in Figure \ref{fig:all_traj}. The estimated likelihood, information criteria (including AIC and BIC), and the residual variance of each parallel growth curve model are provided in Table \ref{tbl:info_RM}. The full PBLSGM for joint development of reading ability and mathematics ability was the unequivocally `best' one, though parametric nonlinear trajectories such as quadratic or Jenss-Bayley fit better than the bilinear spline functional form for the univariate development of mathematics ability as shown in Table \ref{tbl:info_uni}.

\figurehere{4}

\tablehere{9}

Table \ref{tbl:est_RM} presents the estimates of parameters of interest for the joint development of reading and mathematics ability. Post-knot development in reading skills and mathematics skills slowed down substantially. On average, for reading ability, the development rates were $2.030$ and $0.678$ per month in the pre- and post-knot stage, respectively. These rates of mathematics ability were $1.787$ and $0.737$. The transition to the slower growth rate occurred earlier in reading ability ($95$ months) than in mathematics ability ($100$ months) on average, and the age at which such transition occurred was individually different. The estimated knot standard deviation of reading ability and mathematics ability was $3.261$ ($\sqrt{10.632}$) and $4.125$ ($\sqrt{17.019}$) months. It tells us that, for reading (mathematics) ability, the individual students had `personal' knots that were $3.261$ ($4.125$) months earlier or later than the average knot about $68\%$ of the time, and up to $6.522$ ($8.250$) months earlier or later about $95\%$ of the time. There also was a positive association between the development reading skills and mathematics skills indicated by statistically significant intercept-intercept, pre-knot slope-slope, and knot-knot covariances. 

\tablehere{10}

Standardizing the covariances, the intercept-intercept, pre-knot slope-slope, post-knot slope-slope and knot-knot correlations were $0.75$, $0.64$, $0.34$ and $0.59$, respectively. It suggests that, on average, a child who was higher in reading ability tended to be higher in mathematics ability and vice versa. Additionally, on average, a child who was increasing more rapidly in reading ability tended to increase more rapidly in mathematics ability over time and vice versa. Moreover, a child who achieved the change-point of reading ability development earlier tended to arrive at the knot of mathematics ability development earlier and vice versa. 

\subsubsection*{Sensitivity Analysis}
We built PBLSGMs with the grade-in-school as the time scale as a sensitivity analysis. The estimated likelihood, AIC, BIC, and residual variances of the full PBLSGM and its reduced form are also listed in Table \ref{tbl:info_RM}. Note that the full model still performed better than the reduced model as its estimated likelihood was larger, and the BIC/AIC was smaller\footnote{Note that the models constructed for the sensitivity analysis cannot be compared with those for the main analysis directly as the raw datasets used in the two parts were different.}. Upon further investigation, the development in reading skills and mathematics skills slowed down between Wave $5$ and $6$ (around the spring semester of Grade $2$) and between Wave $6$ and $7$ (around the spring semester of Grade $3$), respectively. The estimated knot locations were matched with those from the main analysis: students in Grade $2$ are usually $7$ to $8$ years old, and those in Grade $3$ are usually $8$ to $9$ years old. The knots-of-grade were more heterogeneous than the knots-of-age: the estimated knot standard deviations of reading ability development and mathematics ability development were $4.165$ and $4.817$, respectively. One possible explanation for the discrepancy in terms of the knot variability is that reading/mathematics ability development is more sensitive to changes in age than those in the measurement time and that students in the same grade-in-school could be of different ages.

\subsection{Joint Development in Mathematics Ability and Science Ability}
To demonstrate more complex parallel nonlinear change patterns in practice, we also built PBLSGMs to assess the joint development of mathematics and science ability. It is noted that the mathematics ability was assessed in nine waves, while science ability was only evaluated in eight waves (starting from the second round). The proposed PBLSGMs are capable of addressing this issue since the outcome-specific growth factors are indicated by their own set of repeated measurements in model specification. For this analysis, we defined nine factor loadings for each child from individual measurement occasions of nine waves and specified all of them to the mathematics scores but only the last eight loadings, which corresponded to the second round to the ninth round measurement, to science scores. We fit the full PBLSGM and its reduced version and summarized the estimated likelihood, information criteria, and residuals of both models in Table \ref{tbl:info_MS}.

From the table, the full PBLSGM has a larger estimated likelihood, smaller AIC, and smaller residuals, but its BIC is larger than the corresponding value of its reduced version, suggesting that the fit information fails to lead unequivocal selection. Upon further investigation, the estimated knot variance of science ability was not significant, although it was in the univariate analysis. We then fit a mixed PBLSGM, where the change-point of mathematics ability development was individually different while that of science ability development was assumed to be the same. Its estimated likelihood and information criteria are also listed in Table \ref{tbl:info_MS}. The mixed PBLSGM was the `best' model among three the parallel bilinear spline growth models determined by the AIC or BIC.  

\tablehere{11}

Table \ref{tbl:est_MS} presents the estimates of parameters of interest for the joint development of mathematics and science ability. The estimated trajectory of mathematics ability was identical as the trajectory obtained in the parallel growth model of reading and mathematics IRT scores. Additionally, we noticed that post-knot development in science skills also slowed down. On average, the development rates were $0.839$ and $0.575$ per month in the pre- and post-knot stage. The estimated knot of science ability was at $100$ months (also around $8$ years old) on average. The intercept-intercept, pre-knot slope-slope and post-knot slope-slope correlations were $0.65$, $0.60$ and $0.26$, respectively. It suggests that the initial status of mathematics and science ability and the development rates of these two skills were positively associated.

\tablehere{12}

\section{Discussion}\label{discussion}
In this article, we presented a PBLSGM for assessing nonlinear joint development. With this model, we are capable of estimating the knots, knot variances and knot-knot association. We also proposed its reduced version, as a parsimonious backup for the situations that the full model fails to converge or generates improper solutions, or that estimating knot variances are not of research interest. More importantly, we extended (inverse-) transformation function and matrix in an existing study to the PBLSGM framework to help select proper initial values and, in turn, accelerate the computational process as well as make statistical inferences and interpret the estimates that are directly related to the underlying change patterns. Through simulation studies, for both PBLSGMs, we performed in-depth investigations in terms of the convergence rate, proportion of improper solutions, and performance measures, including the relative bias, empirical standard errors, relative RMSE and coverage probability of each parameter of interest. We also illustrated the proposed models using an empirical data set from a longitudinal study of reading, mathematics and science ability. The results demonstrate the model's valuable capabilities of estimating the knots, their variances and covariance in the framework of individual measurement occasions and interpreting the estimates in the original parameter setting. 

Across all conditions in the simulation design, the convergence rate of the full PBLSGM achieved at least $95\%$, while that of the reduced model was $100\%$. Additionally, as shown in the application section, the full PBLSGM arrived at the convergence status without computational burdens. The full model may suffer an issue of improper solutions, including negative knot variances and out-of-range knot correlations with other growth factors, within- or between-constructs. It is not surprising since the knot variances were set at a moderate level, $0.3$, in the simulation design. In terms of performance measures, the full PBLSGM was capable of estimating the means of the growth factors unbiasedly, precisely, and exhibiting appropriate empirical coverage of nominal $95\%$ confidence intervals.  Additionally, the estimates of the variances and covariances of growth factors often performed well. However, in cases with small sample sizes and/or less precise measurements these estimates exhibited some bias greater than $10\%$. The reduced model produced slightly biased estimates, but its empirical standard errors were comparable to those from the full model. Accordingly, we recommend implementing the reduced model when the full version fails to converge or generates improper solutions at a little cost of a small increase in bias.

We also illustrated how to apply the proposed models on a subset with $n=400$ from ECLS-K: 2011, demonstrating the procedure and providing a set of recommendations for possible issues that empirical researchers may face in practice. First, proper research questions need to be raised before conducting any analysis. These questions include whether or not to test associations between two endpoints as well as their nonlinear change patterns, and for each repeated outcome, whether the research interest lies in estimating a knot and its variance or fitting nonlinear trajectories. It is worth considering either the full or reduced or mixed PBLSGM to estimate knots for a joint developmental process and assess the associations between development rates of different stages if it is the research interest. If the interest lies in fitting multivariate nonlinear developmental processes, it may be appropriate to fit the parallel change patterns with several functional forms and select `best' one. If deciding to fit a PBLSGM, we recommend building univariate bilinear spline growth models first at the reviewers' advice. The univariate growth models allow us to examine whether the linear-linear functional form is appropriate for the underlying change patterns and whether random effects of growth factors exist for constructing a PBLSGM to analyze the joint development. 

We still recommend fitting a pool of candidate parallel growth models with different functional forms and conducting model selection as the `best' functional form from the univariate and parallel growth model could be different. As shown in the first empirical study, the quadratic functional form is the `best' for the univariate development in mathematics ability; however, the PBLSGM with random knots is the `best' for analyzing the joint development of reading and mathematics ability. Though the statistical criteria, such as the estimated likelihood, AIC and BIC, led to unequivocal selection in the analysis of the joint development of reading and mathematics ability, it is not always the case. Other criteria, such as the fit between the model-implied curve and the smooth line of the observed repeated outcome, also helps make a decision. In the first empirical case, for example, the PBLSGMs are better if it is important to capture children's reading ability and mathematics ability in the early stage of the study. 

As advised by reviewers, we conducted a sensitivity analysis in the first empirical example where we use the grade-in-school as the time scale to examine whether the time metric affects the data analysis results. The estimates of the fixed effects of the knots from the models with different time metrics were matched, although the knots-of-grade were more heterogeneous than the knots-of-age. This was not unexpected as students in the same grade-in-school could be of different ages. Accordingly, in an empirical analysis, the selection of the time metric should be, again, driven by research questions instead of any statistical criteria. For example, the structured measurement time and the framework of individual measurement occasions may be considered if the research interest lies in examining the development by grade-in-school and by age when analyzing the longitudinal data of mathematics IRT scores. 

A variety of statistical models have been proposed to investigate changes in multiple endpoints simultaneously. Another possible way is to estimate the effect a time-varying covariate has on the growth trajectory of the other endpoint(s). Additionally, when analyzing parallel growth curves, cross-construct growth factor relationships can also be unidirectional so that we can estimate regression coefficients instead of covariances between growth factors. In this article, we focused on the bivariate growth curve model with a bilinear spline functional form because it is the most straightforward but useful. The proposed model allows for several extensions. First, the functional form of trajectories of each endpoint can be generalized to a linear spline with multiple knots or a nonlinear spline (such as a linear-polynomial or a polynomial-polynomial piecewise), which may also prove useful in real data analyses. Second, though constructed in the framework of individual measurement occasions, the developed PBLSGM considers the same time structure for both repeated outcomes. However, it is possible to be extended for a joint developmental process with varying time structure of each endpoint thanks to the definition variables approach. Note that the same number of repeated measures is not necessary as what is demonstrated for the analysis of the joint development of mathematics and science ability. Third, it is also possible to extend the current work to address a bivariate longitudinal study with dropout under the assumption of missing at random due to the FIML technique. Additionally, we can also extend the PBLSGM to investigate a joint development with at least three constructs. The \textit{OpenMx} and \textit{Mplus} 8 syntax that we provide in the online appendix can also be extended accordingly.

\bibliographystyle{apalike}
\bibliography{Extension1}

\appendix
\renewcommand{\theequation}{A.\arabic{equation}}
\setcounter{equation}{0}

\renewcommand{\thesection}{Appendix \Alph{section}}
\renewcommand{\thesubsection}{A.\arabic{subsection}}

\section{{\textbf{Formula Derivation}}}
\subsection{\textbf{The Reparameterizing Procedure for outcome-specific Growth Factors}}\label{Supp:1A}
For each individual, we have four growth factors to determine the underlying functional form of repeated measurements of $y_{ij}$ in the original setting of a bilinear spline model: the intercept (i.e., the measurement at $t_{0}$, $\eta^{[y]}_{0i}$), one slope for each stage ($\eta^{[y]}_{1i}$ and $\eta^{[y]}_{2i}$, respectively), and the knot ($\gamma^{[y]}_{i}$). To estimate the knot, we may reparameterize the first three individual-level growth factors as the measurement at the knot (i.e., $\eta^{[y]}_{0i}+\eta^{[y]}_{1i}\gamma^{[y]}_{i}$), the mean of two slopes (i.e., $\frac{\eta^{[y]}_{1i}+\eta^{[y]}_{2i}}{2}$), and the half difference between two slopes (i.e., $\frac{\eta^{[y]}_{2i}-\eta^{[y]}_{1i}}{2}$) for the $i^{th}$ individual \citep{Seber2003nonlinear}.

\figurehere{A.1}

\citet{Tishler1981nonlinear} and \citet{Seber2003nonlinear} have proved that a linear-linear regression model can be expressed as either the maximum or minimum response value of two trajectories. \citet{Liu2019knot} and \citet{Liu2019BLSGM} extended such expressions to the framework of BLSGM and showed that there were two possible forms of bilinear spline for the $i^{th}$ individual as shown in Figure \ref{fig:proj1_2cases}. In the left panel ($\eta_{1i}^{[y]}>\eta_{2i}^{[y]}$), the measurement $y_{ij}$ is always the minimum value of two lines and $y_{ij}=\min{(\eta^{[y]}_{0i}+\eta^{[y]}_{1i}t_{ij}, \eta^{[y]}_{02i}+\eta^{[y]}_{2i}t_{ij})}$. The measurements pre- and post-knot can be unified
\begin{equation}\label{eq:left}
\begin{aligned}
y_{ij} &= \min{(\eta^{[y]}_{0i} + \eta^{[y]}_{1i}t_{ij}, \eta^{[y]}_{02i} + \eta^{[y]}_{2i}t_{ij})}\\
&= \frac{1}{2}\big(\eta^{[y]}_{0i} + \eta^{[y]}_{1i}t_{ij} + \eta^{[y]}_{02i} + \eta^{[y]}_{2i}t_{ij} - 
|\eta^{[y]}_{0i} + \eta^{[y]}_{1i}t_{ij} - \eta^{[y]}_{02i} - \eta^{[y]}_{2i}t_{ij}|\big)\\
&= \frac{1}{2}\big(\eta^{[y]}_{0i} + \eta^{[y]}_{1i}t_{ij} + \eta^{[y]}_{02i} + \eta^{[y]}_{2i}t_{ij}\big) - 
\frac{1}{2}\big(|\eta^{[y]}_{0i} + \eta^{[y]}_{1i}t_{ij} - \eta^{[y]}_{02i} - \eta^{[y]}_{2i}t_{ij}|\big)\\
&= \frac{1}{2}\big(\eta^{[y]}_{0i} + \eta^{[y]}_{02i} + \eta^{[y]}_{1i}t_{ij} + \eta^{[y]}_{2i}t_{ij}\big) - 
\frac{1}{2}\big(\eta^{[y]}_{1i} - \eta^{[y]}_{2i}\big)|t_{ij} - \gamma^{[y]}_{i}|\\
&= \eta_{0i}^{'[y]} + \eta_{1i}^{'[y]}\big(t_{ij}-\gamma^{[y]}_{i}\big) + \eta_{2i}^{'[y]}|t_{ij} - \gamma^{[y]}_{i}|\\
&= \eta_{0i}^{'[y]} + \eta_{1i}^{'[y]}\big(t_{ij}-\gamma^{[y]}_{i}\big) + \eta_{2i}^{'[y]}\sqrt{(t_{ij} - \gamma^{[y]}_{i})^2},
\end{aligned}
\end{equation}
where $\eta_{0i}^{'[y]}$, $\eta_{1i}^{'[y]}$ and $\eta_{2i}^{'[y]}$ are the measurement at the knot, the mean of two slopes, and the half difference between two slopes of the trajectory of $y_{ij}$. With straightforward algebra, the outcome $y_{ij}$ of the bilinear spline in the right panel, where the measurement $y_{ij}$ is always the maximum value of two lines, has the same final expression as shown in Equation \ref{eq:left}. By applying such transformation to each repeated outcome, we obtain the outcome-specific reparameterized growth factors. 

\subsection{Taylor Series Expansion}\label{Supp:1B}
Following \citet{Liu2019BLSGM}, for the $i^{th}$ individual, we write a repeated outcome as a function of its trajectory knot, $f(\gamma^{[y]}_{i})$ and obtain its first derivative with respect to the knot
\begin{equation}\nonumber
f(\gamma^{[y]}_{i})=\eta_{0i}^{'[y]} + \eta_{1i}^{'[y]}\big(t_{ij}-\gamma^{[y]}_{i}\big) + \eta_{2i}^{'[y]}\sqrt{(t_{ij} - \gamma^{[y]}_{i})^2} 
\end{equation}
and
\begin{equation}\nonumber
f^{'}(\gamma^{[y]}_{i})=\eta^{[y]}_{1i}-\eta_{1i}^{'[y]}-\frac{\eta_{2i}^{'[y]}(t_{ij}-\gamma^{[y]}_{i})}{\sqrt{(t_{ij}-\gamma^{[y]}_{i})^2}}=-\eta_{2i}^{'[y]}-\frac{\eta_{2i}^{'[y]}(t_{ij}-\gamma^{[y]}_{i})}{\sqrt{(t_{ij}-\gamma^{[y]}_{i})^2}},
\end{equation}
respectively. Then for $f(\gamma^{[y]}_{i})$, we conducted the Taylor series expansion and expressed it as 
\begin{equation}\nonumber
\begin{aligned}
f(\gamma^{[y]}_{i})&= f(\mu^{[y]}_{\gamma})+\frac{f'(\mu^{[y]}_{\gamma})}{1!}(\gamma^{[y]}_{i}-\mu^{[y]}_{\gamma})+\cdots\\
&= \eta_{0i}^{'[y]}+\eta_{1i}^{'[y]}(t_{ij}-\mu^{[y]}_{\gamma})+\eta_{2i}^{'[y]}\sqrt{(t_{ij} - \mu^{[y]}_{\gamma})^2} +
(\gamma^{[y]}_{i}-\mu^{[y]}_{\gamma})\bigg[-\eta_{2i}^{'[y]}-\frac{\eta_{2i}^{'[y]}(t_{ij}-\mu^{[y]}_{\gamma})}{|t_{ij}-\mu^{[y]}_{\gamma}|}\bigg]+\cdots\\
&\approx \eta_{0i}^{'[y]}+\eta_{1i}^{'[y]}(t_{ij}-\mu^{[y]}_{\gamma})+\eta_{2i}^{'[y]}|t_{ij}-\mu^{[y]}_{\gamma}| +
(\gamma^{[y]}_{i}-\mu^{[y]}_{\gamma})\bigg[-\mu^{'[y]}_{\eta_{2}}-\frac{\mu^{'[y]}_{\eta_{2}}(t_{ij}-\mu^{[y]}_{\gamma})}{|t_{ij}-\mu^{[y]}_{\gamma}|}\bigg],
\end{aligned}
\end{equation}
from which we then have the individual-level reparameterized growth factors and their factor loadings.

\newpage

\subsection{\textbf{Transformation between Two Parameter-spaces}}\label{Supp:1C}
\subsubsection{Transformation Functions and Matrices}
Suppose $\boldsymbol{f}: \mathcal{R}^{4}\rightarrow \mathcal{R}^{4}$ is a function, which takes a point $\boldsymbol{\eta}_{i}^{[u]}\in\mathcal{R}^{4}$ as input and produces the vector $\boldsymbol{f}(\boldsymbol{\eta}_{i}^{[u]})\in\mathcal{R}^{4}$ (i.e., $\boldsymbol{\eta}_{i}^{'[u]}\in\mathcal{R}^{4}$) as output. By the multivariate delta method \citep{Lehmann1998Delta},
\begin{equation}\label{eq:trans_fun}
\boldsymbol{\eta}_{i}^{'[u]}=\boldsymbol{f}(\boldsymbol{\eta}_{i}^{[u]})\sim N\bigg(\boldsymbol{f}(\boldsymbol{\mu}^{[u]}_{\boldsymbol{\eta}}), \boldsymbol{\nabla}_{\boldsymbol{f}}(\boldsymbol{\mu}^{[u]}_{\boldsymbol{\eta}})\boldsymbol{\Psi}^{[u]}_{\boldsymbol{\eta}}\boldsymbol{\nabla}^{T}_{\boldsymbol{f}}(\boldsymbol{\mu}^{[u]}_{\boldsymbol{\eta}})\bigg), 
\end{equation}
where $\boldsymbol{\mu}^{[u]}_{\boldsymbol{\eta}}$ and $\boldsymbol{\Psi}^{[u]}_{\boldsymbol{\eta}}$ are the mean vector and variance-covariance matrix of the outcome-specific growth factors in the original setting, and $\boldsymbol{f}$ is defined as
\begin{equation}\nonumber
\boldsymbol{f}(\boldsymbol{\eta}_{i}^{[u]})=\left(\begin{array}{rrrr}
\eta^{[u]}_{0i}+\gamma^{[u]}_{i}\eta^{[u]}_{1i} & \frac{\eta^{[u]}_{1i}+\eta^{[u]}_{2i}}{2} & \frac{\eta^{[u]}_{2i}-\eta^{[u]}_{1i}}{2} &
\gamma^{[u]}_{i}-\mu^{[u]}_{\gamma}
\end{array}\right)^{T}.
\end{equation}

Similarly, suppose $\boldsymbol{h}: \mathcal{R}^{4}\rightarrow \mathcal{R}^{4}$ is a function, which takes a point $\boldsymbol{\eta}_{i}^{'[u]}\in\mathcal{R}^{4}$ as input and produces the vector $\boldsymbol{h}(\boldsymbol{\eta}_{i}^{'[u]})\in\mathcal{R}^{4}$ (i.e., $\boldsymbol{\eta}_{i}^{[u]}\in\mathcal{R}^{4}$) as output. By the multivariate delta method,
\begin{equation}\label{eq:inverse_fun}
\boldsymbol{\eta}_{i}^{[u]}=\boldsymbol{h}(\boldsymbol{\eta}_{i}^{'[u]})\sim N\bigg(\boldsymbol{h}(\boldsymbol{\mu}^{'[u]}_{\boldsymbol{\eta}}), \boldsymbol{\nabla}_{\boldsymbol{h}}(\boldsymbol{\mu}^{'[u]}_{\boldsymbol{\eta}})\boldsymbol{\Psi}^{'[u]}_{\boldsymbol{\eta}}\boldsymbol{\nabla}^{T}_{\boldsymbol{h}}(\boldsymbol{\mu}^{'[u]}_{\boldsymbol{\eta}})\bigg), 
\end{equation}
where $\boldsymbol{\mu}^{'[u]}_{\boldsymbol{\eta}}$ and $\boldsymbol{\Psi}^{'[u]}_{\boldsymbol{\eta}}$ are the mean vector and variance-covariance matrix of the outcome-specific growth factors in the reparameterized frame, and $\boldsymbol{h}$ is defined as
\begin{equation}\nonumber
\boldsymbol{h}(\boldsymbol{\eta}_{i}^{'[u]})=\left(\begin{array}{rrrr}
\eta^{'[u]}_{0i}-\gamma_{i}^{[u]}\eta^{'[u]}_{1i}+\gamma_{i}^{[u]}\eta^{'[u]}_{2i} & \eta^{'[u]}_{1i}-\eta^{'[u]}_{2i} & \eta^{'[u]}_{1i}+\eta^{'[u]}_{2i} & \delta_{i}^{[u]}+\mu^{[u]}_{\gamma}
\end{array}\right)^{T}.
\end{equation}

Based on Equations (\ref{eq:trans_fun}) and (\ref{eq:inverse_fun}), we can make the transformation between the growth factor means of two parameter spaces by $\boldsymbol{\mu}^{'[u]}_{\boldsymbol{\eta}}
\approx\boldsymbol{f}(\boldsymbol{\mu}^{[u]}_{\boldsymbol{\eta}})$ and $\boldsymbol{\mu}^{[u]}_{\boldsymbol{\eta}}
\approx\boldsymbol{h}(\boldsymbol{\mu}^{'[u]}_{\boldsymbol{\eta}})$, respectively. We then express the transformation and inverse-transformation matrix between the variance-covariance matrix of the growth factors in the original setting and that in the reparameterized frame as
\begin{equation}\nonumber
\begin{aligned}
\begin{pmatrix}
\boldsymbol{\Psi}^{'[y]}_{\boldsymbol{\eta}} & \boldsymbol{\Psi}^{'[yz]}_{\boldsymbol{\eta}} \\
& \boldsymbol{\Psi}^{'[z]}_{\boldsymbol{\eta}}
\end{pmatrix}
&\approx\begin{pmatrix}
\boldsymbol{\nabla}_{\boldsymbol{f}}(\boldsymbol{\mu}^{[y]}_{\boldsymbol{\eta}}) & \boldsymbol{0} \\
& \boldsymbol{\nabla}_{\boldsymbol{f}}(\boldsymbol{\mu}^{[z]}_{\boldsymbol{\eta}})
\end{pmatrix}
\times\begin{pmatrix}
\boldsymbol{\Psi}^{[y]}_{\boldsymbol{\eta}} & \boldsymbol{\Psi}^{[yz]}_{\boldsymbol{\eta}} \\
& \boldsymbol{\Psi}^{[z]}_{\boldsymbol{\eta}}
\end{pmatrix}
\times\begin{pmatrix}
\boldsymbol{\nabla}_{\boldsymbol{f}}(\boldsymbol{\mu}^{[y]}_{\boldsymbol{\eta}}) & \boldsymbol{0} \\
& \boldsymbol{\nabla}_{\boldsymbol{f}}(\boldsymbol{\mu}^{[z]}_{\boldsymbol{\eta}})
\end{pmatrix}^{T}\ \ \ \ \ \ \ \ \ \ \ \ \ \ \ \ \ \ \ \ \\
&=\begin{pmatrix}
\boldsymbol{\nabla}_{\boldsymbol{f}}(\boldsymbol{\mu}^{[y]}_{\boldsymbol{\eta}})\times\boldsymbol{\Psi}^{[y]}_{\boldsymbol{\eta}}\times\boldsymbol{\nabla}^{T}_{\boldsymbol{f}}(\boldsymbol{\mu}^{[y]}_{\boldsymbol{\eta}}) &
\boldsymbol{\nabla}_{\boldsymbol{f}}(\boldsymbol{\mu}^{[y]}_{\boldsymbol{\eta}})\times\boldsymbol{\Psi}^{[yz]}_{\boldsymbol{\eta}}\times\boldsymbol{\nabla}^{T}_{\boldsymbol{f}}(\boldsymbol{\mu}^{[z]}_{\boldsymbol{\eta}}) \\
\boldsymbol{\nabla}_{\boldsymbol{f}}(\boldsymbol{\mu}^{[z]}_{\boldsymbol{\eta}})\times\boldsymbol{\Psi}^{[yz]}_{\boldsymbol{\eta}}\times\boldsymbol{\nabla}^{T}_{\boldsymbol{f}}(\boldsymbol{\mu}^{[y]}_{\boldsymbol{\eta}}) &
\boldsymbol{\nabla}_{\boldsymbol{f}}(\boldsymbol{\mu}^{[z]}_{\boldsymbol{\eta}})\times\boldsymbol{\Psi}^{[z]}_{\boldsymbol{\eta}}\times\boldsymbol{\nabla}^{T}_{\boldsymbol{f}}(\boldsymbol{\mu}^{[z]}_{\boldsymbol{\eta}})
\end{pmatrix}
\end{aligned}
\end{equation}
and
\begin{equation}\nonumber
\begin{aligned}
\begin{pmatrix}
\boldsymbol{\Psi}^{[y]}_{\boldsymbol{\eta}} & \boldsymbol{\Psi}^{[yz]}_{\boldsymbol{\eta}} \\
& \boldsymbol{\Psi}^{[z]}_{\boldsymbol{\eta}}
\end{pmatrix}
&\approx\begin{pmatrix}
\boldsymbol{\nabla}_{\boldsymbol{h}}(\boldsymbol{\mu}^{'[y]}_{\boldsymbol{\eta}}) & \boldsymbol{0} \\
& \boldsymbol{\nabla}_{\boldsymbol{h}}(\boldsymbol{\mu}^{'[z]}_{\boldsymbol{\eta}})
\end{pmatrix}
\times
\begin{pmatrix}
\boldsymbol{\Psi}^{'[y]}_{\boldsymbol{\eta}} & \boldsymbol{\Psi}^{'[yz]}_{\boldsymbol{\eta}} \\
& \boldsymbol{\Psi}^{'[z]}_{\boldsymbol{\eta}}
\end{pmatrix}
\times\begin{pmatrix}
\boldsymbol{\nabla}_{\boldsymbol{h}}(\boldsymbol{\mu}^{'[y]}_{\boldsymbol{\eta}}) & \boldsymbol{0} \\
& \boldsymbol{\nabla}_{\boldsymbol{h}}(\boldsymbol{\mu}^{'[z]}_{\boldsymbol{\eta}})
\end{pmatrix}^{T}\\
&=\begin{pmatrix}
\boldsymbol{\nabla}_{\boldsymbol{h}}(\boldsymbol{\mu}^{'[y]}_{\boldsymbol{\eta}})\times\boldsymbol{\Psi}^{'[y]}_{\boldsymbol{\eta}}\times\boldsymbol{\nabla}^{T}_{\boldsymbol{h}}(\boldsymbol{\mu}^{'[y]}_{\boldsymbol{\eta}}) &
\boldsymbol{\nabla}_{\boldsymbol{h}}(\boldsymbol{\mu}^{'[y]}_{\boldsymbol{\eta}})\times\boldsymbol{\Psi}^{'[yz]}_{\boldsymbol{\eta}}\times\boldsymbol{\nabla}^{T}_{\boldsymbol{h}}(\boldsymbol{\mu}^{'[z]}_{\boldsymbol{\eta}}) \\
\boldsymbol{\nabla}_{\boldsymbol{h}}(\boldsymbol{\mu}^{'[z]}_{\boldsymbol{\eta}})\times\boldsymbol{\Psi}^{'[yz]}_{\boldsymbol{\eta}}\times\boldsymbol{\nabla}^{T}_{\boldsymbol{h}}(\boldsymbol{\mu}^{'[y]}_{\boldsymbol{\eta}}) &
\boldsymbol{\nabla}_{\boldsymbol{h}}(\boldsymbol{\mu}^{'[z]}_{\boldsymbol{\eta}})\times\boldsymbol{\Psi}^{'[z]}_{\boldsymbol{\eta}}\times\boldsymbol{\nabla}^{T}_{\boldsymbol{h}}(\boldsymbol{\mu}^{'[z]}_{\boldsymbol{\eta}})
\end{pmatrix},
\end{aligned}
\end{equation}
respectively. In above two equations, $\boldsymbol{\Psi}^{[u]}_{\boldsymbol{\eta}}$  ($\boldsymbol{\Psi}^{'[u]}_{\boldsymbol{\eta}}$) $(u=y,z)$ and  $\boldsymbol{\Psi}^{[yz]}_{\boldsymbol{\eta}}$ ($\boldsymbol{\Psi}^{'[yz]}_{\boldsymbol{\eta}}$) are $4\times 4$ outcome-specific variance-covariance matrix of original (reparameterized) growth factors and the covariances between growth factors of the bivariate repeated outcomes in the original (reparameterized) framework, respectively. Additionally, $\boldsymbol{\nabla}_{\boldsymbol{f}}(\boldsymbol{\mu}^{[u]}_{\boldsymbol{\eta}})$ and $\boldsymbol{\nabla}_{\boldsymbol{h}}(\boldsymbol{\mu}^{'[u]}_{\boldsymbol{\eta}})$ are defined as
\begin{equation}\nonumber
\boldsymbol{\nabla}_{\boldsymbol{f}}(\boldsymbol{\mu}^{[u]}_{\boldsymbol{\eta}})
=\left(\begin{array}{rrrr}
1 & \mu^{[u]}_{\gamma} & 0 & \mu^{[u]}_{\eta_{1}}\\
0 & 0.5 & 0.5 & 0\\
0 & -0.5 & 0.5 & 0\\
0 & 0 & 0 & 1 
\end{array}\right)
\end{equation}
and
\begin{equation}\nonumber
\boldsymbol{\nabla}_{\boldsymbol{h}}(\boldsymbol{\mu}^{'[u]}_{\boldsymbol{\eta}})=\left(\begin{array}{rrrr}
1 & -\mu^{[u]}_{\gamma} & \mu^{[u]}_{\gamma} & 0\\0 & 1 & -1 & 0\\0 & 1 & 1 & 0\\0 & 0 & 0 & 1
\end{array}\right),
\end{equation}
respectively.

\subsubsection{Expression of each cell of the re-reparameterized mean vector and variance-covariance matrix}
\begin{equation}
\begin{aligned}
\mu^{[u]}_{\eta_{0}}&\approx\mu^{[u]}_{\eta_{0}^{'}}-\mu^{[u]}_{\gamma}\mu^{[u]}_{\eta_{1}^{'}}+\mu^{[u]}_{\gamma}\mu^{[u]}_{\eta_{2}^{'}}\\
\mu^{[u]}_{\eta_{1}}&=\mu^{[u]}_{\eta_{1}^{'}}-\mu^{[u]}_{\eta_{2}^{'}}\\
\mu^{[u]}_{\eta_{2}}&=\mu^{[u]}_{\eta_{2}^{'}}+\mu^{[u]}_{\eta_{1}^{'}}\\
\mu^{[u]}_{\gamma}&=\mu^{[u]}_{\gamma}\\
\psi^{[u]}_{00}&\approx(\psi_{11}^{'[u]}+\psi_{22}^{'[u]}-2\psi_{12}^{'[u]})\mu_{\gamma}^{[u]2}+2(\psi_{02}^{'[u]}-\psi_{01}^{'[u]})\mu^{[u]}_{\gamma}+\psi_{00}^{'[u]}\nonumber\\
\psi^{[u]}_{01}&\approx(2\psi_{12}^{'[u]}-\psi_{11}^{'[u]}-\psi_{22}^{'[u]})\mu^{[u]}_{\gamma}+(\psi_{01}^{'[u]}-\psi_{02}^{'[u]})\nonumber\\
\psi^{[u]}_{02}&\approx(\psi_{22}^{'[u]}-\psi_{11}^{'[u]})\mu^{[u]}_{\gamma}+(\psi_{01}^{'[u]}+\psi_{02}^{'[u]})\nonumber\\
\psi^{[u]}_{0\gamma}&\approx(\psi_{2\gamma}^{'[u]}-\psi_{1\gamma}^{'[u]})\mu^{[u]}_{\gamma}+\psi_{0\gamma}^{'[u]}\nonumber\\
\psi^{[u]}_{11}&=\psi_{11}^{'[u]}+\psi_{22}^{'[u]}-2\psi_{12}^{'[u]}\nonumber\\
\psi^{[u]}_{12}&=\psi_{11}^{'[u]}-\psi_{22}^{'[u]}\nonumber\\
\psi^{[u]}_{1\gamma}&=\psi_{1\gamma}^{'[u]}-\psi_{2\gamma}^{'[u]}\nonumber\\
\psi^{[u]}_{22}&=\psi_{11}^{'[u]}+\psi_{22}^{'[u]}+2\psi_{12}^{'[u]}\nonumber\\
\psi^{[u]}_{2\gamma}&=\psi_{1\gamma}^{'[u]}+\psi_{2\gamma}^{'[u]}\nonumber\\
\psi^{[u]}_{\gamma\gamma}&=\psi_{\gamma\gamma}^{'[u]}\nonumber\\
\psi^{[yz]}_{00}&\approx(\psi_{11}^{'[yz]}+\psi_{22}^{'[yz]}-\psi_{12}^{'[yz]}-\psi_{21}^{'[yz]})\mu^{[y]}_{\gamma}\mu^{[z]}_{\gamma}+(\psi_{20}^{'[yz]}-\psi_{10}^{'[yz]})\mu^{[y]}_{\gamma}+(\psi_{02}^{'[yz]}-\psi_{01}^{'[yz]})\mu^{[z]}_{\gamma}+\psi_{00}^{'[yz]}\nonumber\\
\psi^{[yz]}_{01}&\approx(\psi_{12}^{'[yz]}+\psi_{21}^{'[yz]}-\psi_{11}^{'[yz]}-\psi_{22}^{'[yz]})\mu^{[y]}_{\gamma}+(\psi_{01}^{'[yz]}-\psi_{02}^{'[yz]})\nonumber\\
\psi^{[yz]}_{02}&\approx(\psi_{21}^{'[yz]}-\psi_{12}^{'[yz]}-\psi_{11}^{'[yz]}+\psi_{22}^{'[yz]})\mu^{[y]}_{\gamma}+(\psi_{01}^{'[yz]}+\psi_{02}^{'[yz]})\nonumber\\
\psi^{[yz]}_{0\gamma}&\approx\psi_{0\gamma}^{'[yz]}+(\psi_{2\gamma}^{'[yz]}-\psi_{1\gamma}^{'[yz]})\mu^{[y]}_{\gamma}\nonumber\\
\psi^{[yz]}_{10}&\approx(\psi_{12}^{'[yz]}+\psi_{21}^{'[yz]}-\psi_{11}^{'[yz]}-\psi_{22}^{'[yz]})\mu^{[z]}_{\gamma}+(\psi_{10}^{'[yz]}-\psi_{20}^{'[yz]})\nonumber\\
\psi^{[yz]}_{11}&=\psi_{11}^{'[yz]}-\psi_{21}^{'[yz]}-\psi_{12}^{'[yz]}+\psi_{22}^{'[yz]}\nonumber\\
\psi^{[yz]}_{12}&=\psi_{11}^{'[yz]}-\psi_{21}^{'[yz]}+\psi_{12}^{'[yz]}-\psi_{22}^{'[yz]}\nonumber\\
\psi^{[yz]}_{1\gamma}&=\psi_{1\gamma}^{'[yz]}-\psi_{2\gamma}^{'[yz]}\nonumber\\
\psi^{[yz]}_{20}&\approx(\psi_{12}^{'[yz]}-\psi_{21}^{'[yz]}-\psi_{11}^{'[yz]}+\psi_{22}^{'[yz]})\mu^{[z]}_{\gamma}+(\psi_{10}^{'[yz]}+\psi_{20}^{'[yz]})\nonumber\\
\psi^{[yz]}_{21}&=\psi_{11}^{'[yz]}-\psi_{22}^{'[yz]}-\psi_{12}^{'[yz]}+\psi_{21}^{'[yz]}\nonumber\\
\psi^{[yz]}_{22}&=\psi_{11}^{'[yz]}+\psi_{22}^{'[yz]}+\psi_{12}^{'[yz]}+\psi_{21}^{'[yz]}\nonumber\\
\psi^{[yz]}_{2\gamma}&=\psi_{1\gamma}^{'[yz]}+\psi_{2\gamma}^{'[yz]}\nonumber\\
\psi^{[yz]}_{\gamma0}&\approx\psi_{\gamma0}^{'[yz]}+(\psi_{\gamma2}^{'[yz]}-\psi_{\gamma1}^{'[yz]})\mu^{[z]}_{\gamma}\nonumber\\
\psi^{[yz]}_{\gamma1}&=\psi_{\gamma1}^{'[yz]}-\psi_{\gamma2}^{'[yz]}\nonumber\\
\psi^{[yz]}_{\gamma2}&=\psi_{\gamma1}^{'[yz]}+\psi_{\gamma2}^{'[yz]}\nonumber\\
\psi^{[yz]}_{\gamma\gamma}&=\psi_{\gamma\gamma}^{'[yz]}\nonumber
\end{aligned}
\end{equation}

\newpage

\section{\textbf{More Results}}\label{Supp:2}
\tablehere{B.1}
\tablehere{B.2}
\tablehere{B.3}
\tablehere{B.4}
\newpage
\newpage
\renewcommand\thetable{\arabic{table}}
\setcounter{table}{0}

\begin{table}[!ht]
\centering
\begin{threeparttable}
\caption{Performance Metrics: Definitions and Estimates}
\begin{tabular}{p{4cm}p{4.5cm}p{5.5cm}}
\hline
\hline
\textbf{Criteria} & \textbf{Definition} & \textbf{Estimate} \\
\hline
Relative Bias & $E_{\hat{\theta}}(\hat{\theta}-\theta)/\theta$ & $\sum_{s=1}^{S}(\hat{\theta}-\theta)/S\theta$ \\
Empirical SE & $\sqrt{Var(\hat{\theta})}$ & $\sqrt{\sum_{s=1}^{S}(\hat{\theta}-\bar{\theta})^{2}/(S-1)}$ \\
Relative RMSE & $\sqrt{E_{\hat{\theta}}(\hat{\theta}-\theta)^{2}}/\theta$ & $\sqrt{\sum_{s=1}^{S}(\hat{\theta}-\theta)^{2}/S}/\theta$ \\
Coverage Probability & $Pr(\hat{\theta}_{\text{low}}\le\theta\le\hat{\theta}_{\text{upper}})$ & $\sum_{s=1}^{S}I(\hat{\theta}_{\text{low},s}\le\theta\le\hat{\theta}_{\text{upper},s})/S$\\
\hline
\hline
\end{tabular}
\label{tbl:metric}
\begin{tablenotes}
\small
\item[1] {$\theta$: the population value of the parameter of interest} \\
\item[2] {$\hat{\theta}$: the estimate of $\theta$} \\
\item[3] {$S$: the number of replications and set as $1,000$ in our simulation study} \\
\item[4] {$s=1,\dots,S$: indexes the replications of the simulation} \\
\item[5] {$\hat{\theta}_{s}$: the estimate of $\theta$ from the $s^{th}$ replication} \\
\item[6] {$\bar{\theta}$: the mean of $\hat{\theta}_{s}$'s across replications} \\
\item[7] {$I()$: an indicator function}
\end{tablenotes}
\end{threeparttable}
\end{table}

\begin{table}[!ht]
\centering
\begin{threeparttable}
\setlength{\tabcolsep}{5pt}
\renewcommand{\arraystretch}{0.75}
\caption{Simulation Design for PBLSGMs with Unknown Knots in the ITPs Framework}
\begin{tabular}{p{7.0cm} p{10.5cm}}
\hline
\hline
\multicolumn{2}{c}{\textbf{Fixed Conditions}} \\
\hline
\textbf{Variables} & \textbf{Conditions} \\
\hline
Intercept Variances & $\psi_{00}^{[u]}=25$ $(u=y,z)$ \\
\hline
Slope Variances & $\psi_{11}^{[u]}=\psi_{22}^{[u]}=1$ $(u=y,z)$ \\
\hline
Knot Variances & $\psi_{\gamma\gamma}^{[u]}=0.09$ $(u=y,z)$ \\
\hline
Correlation of Within-Construct GFs & $\rho^{[u]}=0.3$ $(u=y,z)$ \\
\hline
Residual Correlation & $\rho_{\epsilon}=0.3$ \\
\hline
\hline
\multicolumn{2}{c}{\textbf{Manipulated Conditions}} \\
\hline
\hline
\textbf{Variables} & \textbf{Conditions} \\
\hline
Sample Size & $n=200$ or $500$\\
\hline
\multirow{2}{*}{Time (\textit{t})} & $6$ scaled and equally spaced $t_{j}$ ($j=0$, $\dots$, $J-1$, $J=6$) \\
& $10$ scaled and equally spaced $t_{j}$ ($j=0$, $\dots$, $J-1$, $J=10$) \\
\hline
Individual \textit{t} & $t_{ij} \sim U(t_{j}-\Delta, t_{j}+\Delta)$ ($j=0$, $\dots$, $J-1$; $\Delta=0.25$) \\
\hline
\multirow{3}{*}{Knot Locations} & $\mu_{\gamma}^{[y]}=2.50$; $\mu_{\gamma}^{[z]}=2.50$ for $J=6$\\
& $\mu_{\gamma}^{[y]}=4.50$; $\mu_{\gamma}^{[z]}=4.50$ for $J=10$ \\
& $\mu_{\gamma}^{[y]}=3.50$; $\mu_{\gamma}^{[z]}=5.50$ for $J=10$ \\
\hline
Correlation of Between-Construct GFs & $\rho=-0.3, 0, 0.3$ \\
\hline
Residual Variance & $\theta_{\epsilon}^{[u]}=1$ or $2$ $(u=y,z)$ \\
\hline
\hline
\multicolumn{2}{l}{\textbf{Scenario 1: Different Intercept Mean}} \\
\hline
\textbf{Variables} & \textbf{Conditions} \\
\hline
First Slope Means & $\mu_{\eta_{1}}^{[u]}=5$ $(u=y,z)$\\
\hline
Second Slope Means
& $\mu_{\eta_{2}}^{[u]}=2.6$ $(u=y,z)$ \\
\hline
Intercept Means & $\mu_{\eta_{0}}^{[y]}=98$, $\mu_{\eta_{0}}^{[z]}=102$ \\
\hline
\hline
\multicolumn{2}{l}{\textbf{Scenario 2: Different First Slope Mean}} \\
\hline
\textbf{Variables} & \textbf{Conditions} \\
\hline
Intercept Means & $\mu_{\eta_{0}}^{[u]}=100$ $(u=y,z)$ \\
\hline
Second Slope Means & $\mu_{\eta_{2}}^{[u]}=2$ $(u=y,z)$ \\
\hline
First Slope Means & $\mu_{\eta_{1}}^{[y]}=4.4$, $\mu_{\eta_{1}}^{[z]}=3.6$ \\
\hline
\hline
\multicolumn{2}{l}{\textbf{Scenario 3: Different Second Slope Mean}} \\
\hline
\textbf{Variables} & \textbf{Conditions} \\
\hline
Intercept Means & $\mu_{\eta_{0}}^{[u]}=100$ $(u=y,z)$ \\
\hline
First Slope Means & $\mu_{\eta_{1}}^{[u]}=5$ $(u=y,z)$ \\
\hline
Second Slope Means & $\mu_{\eta_{2}}^{[y]}=2.6$, $\mu_{\eta_{2}}^{[z]}=3.4$ \\
\hline
\hline
\end{tabular}
\label{tbl:simu_design}
\end{threeparttable}
\end{table}

\begin{table}[!ht]
\centering
\begin{threeparttable}
\setlength{\tabcolsep}{5pt}
\renewcommand{\arraystretch}{0.75}
\caption{Number of Improper Solutions among $1,000$ Replications of the PBLSGMs in the ITPs Framework ($10$ Repeated Measurements)}
\begin{tabular}{p{5.0cm}|p{2.5cm}|p{2.0cm}|R{1.5cm}R{1.5cm}|R{1.5cm}R{1.5cm}}
\hline
\hline
& & & \multicolumn{2}{c}{$\theta^{[u]}_{\epsilon}=1$} & \multicolumn{2}{c}{$\theta^{[u]}_{\epsilon}=2$}\\ 
\hline
& & & \textbf{$n=200$} & \textbf{$n=500$} & \textbf{$n=200$} & \textbf{$n=500$}\\
\hline
\hline
\multirow{6}{*}{\makecell[l]{Positive Between-Construct \\ Correlation $\rho=0.3$}}
& \multirow{3}{*}{\makecell[l]{Same Knot\\Locations}} 
& Scenario 1 & $11//55$\tnote{1} & $0//1$ & $188//217$ & $39//92$ \\
& & Scenario 2 & $88//111$ & $5//17$ & $292//277$ & $134//118$ \\
& & Scenario 3 & $105//112$ & $10//17$ & $291//266$ & $122//132$ \\
\cline{2-7}
&\multirow{3}{*}{\makecell[l]{Different Knot \\ Locations}} 
& Scenario 1 & $17//59$ & $1//3$ & $180//261$ & $25//77$ \\
& & Scenario 2 & $70//132$ & $11//21$ & $311//243$ & $133//139$ \\
& & Scenario 3 & $74//120$ & $10//15$ & $288//269$ & $119//147$ \\
\hline
\multirow{6}{*}{\makecell[l]{Negative Between-Construct \\ Correlation $\rho=-0.3$}}
&\multirow{3}{*}{\makecell[l]{Same Knot \\ Locations}} 
& Scenario 1 & $17//100$ & $1//12$ & $172//239$ & $36//124$ \\
& & Scenario 2 & $88//130$ & $10//46$ & $267//269$ & $105//163$ \\
& & Scenario 3 & $73//143$ & $8//45$ & $327//251$ & $135//159$ \\
\cline{2-7}
& \multirow{3}{*}{\makecell[l]{Different Knot \\ Locations}} 
& Scenario 1 & $14//62$ & $0//3$ & $167//232$ & $25//100$ \\
& & Scenario 2 & $77//132$ & $4//26$ & $290//258$ & $114//143$ \\
& & Scenario 3 & $86//139$ & $7//15$ & $295//242$ & $111//154$ \\
\hline
\multirow{6}{*}{\makecell[l]{Zero Between-Construct \\ Correlation $\rho=0$}} 
&\multirow{3}{*}{\makecell[l]{Same Knot \\ Locations}} 
& Scenario 1 & $16//59$ & $0//0$ & $187//209$ & $40//83$ \\
& & Scenario 2 & $76//109$ & $11//15$ & $315//250$ & $100//125$ \\
& & Scenario 3 & $67//114$ & $7//23$ & $294//222$ & $151//129$ \\
\cline{2-7}
&\multirow{3}{*}{\makecell[l]{Different Knot \\ Locations}} 
& Scenario 1 & $13//41$ & $0//1$ & $161//203$ & $36//84$ \\
& & Scenario 2 & $74//75$ & $14//17$ & $311//246$ & $122//145$ \\
& & Scenario 3 & $85//87$ & $9//15$ & $303//223$ & $121//114$ \\
\hline
\hline
\end{tabular}
\label{tbl:Improper}
\begin{tablenotes}
\small
\item[1] {$11//55$ suggests that, for the proposed PBLSGM, among $1,000$ replications with convergent solutions, we have $11$ and $55$ improper solutions result from negative knot variances and out-of-range knot correlations with other growth factors from the same or the other construct, respectively.}
\end{tablenotes}
\end{threeparttable}
\end{table}

\begin{table}[!ht]
\centering
\begin{threeparttable}
\setlength{\tabcolsep}{5pt}
\renewcommand{\arraystretch}{0.75}
\caption{Median and range of the Relative Bias of Each Parameter in PBLSGM in the ITPs Framework  ($10$ Repeated Measurements)}
\begin{tabular}{p{3.0cm}p{1cm}R{5.5cm}R{5.5cm}}
\hline
\hline
& \textbf{Para.} & \textbf{Reduced PBLSGM} & \textbf{Full PBLSGM} \\
\hline
& & Median (Range) & Median (Range) \\
\hline
\hline
\multirow{4}{*}{\textbf{\makecell[l]{Grow Factor \\ Means of Y}}} 
& $\mu^{[y]}_{\eta_{0}}$ & $0.0000$ ($-0.0002$, $0.0004$) & $0.0000$ ($-0.0002$, $0.0003$) \\
& $\mu^{[y]}_{\eta_{1}}$ & $-0.0016$ ($-0.0031$, $0.0000$) & $-0.0014$ ($-0.0026$, $0.0000$) \\
& $\mu^{[y]}_{\eta_{2}}$ & $0.0014$ ($-0.0011$, $0.0040$) & $0.0016$ ($-0.0004$, $0.0040$) \\
& $\mu^{[y]}_{\gamma}$ & $0.0012$ ($-0.0008$, $0.0037$) & $0.0005$ ($-0.0008$, $0.0024$) \\
\hline
\hline
\multirow{4}{*}{\textbf{\makecell[l]{Grow Factor \\ Variances of Y}}} 
& $\psi^{[y]}_{00}$ & $-0.0193$ ($-0.0289$, $-0.0134$) & $-0.0069$ ($-0.0206$, $0.0000$) \\
& $\psi^{[y]}_{11}$ & $0.0856$ ($0.0626$, $0.1278$) & $0.0168$ ($-0.0025$, $0.0712$) \\
& $\psi^{[y]}_{22}$ & $0.0557$ ($0.0351$, $0.0768$) & $0.0092$ ($-0.0027$, $0.0401$) \\
& $\psi^{[y]}_{\gamma\gamma}$ & ---\tnote{1} & $-0.1416$ ($-0.3829$, $0.0309$) \\
\hline
\hline
\multirow{4}{*}{\textbf{\makecell[l]{Grow Factor \\ Means of Z}}} 
& $\mu^{[z]}_{\eta_{0}}$ & $0.0000$ ($-0.0003$, $0.0002$) & $0.0000$ ($-0.0003$, $0.0002$) \\
& $\mu^{[z]}_{\eta_{1}}$ & $-0.0005$ ($-0.0014$, $0.0009$) & $-0.0006$ ($-0.0015$, $0.0007$) \\
& $\mu^{[z]}_{\eta_{2}}$ & $0.0020$ ($-0.0002$, $0.0057$) & $0.0019$ ($-0.0003$, $0.0045$) \\
& $\mu^{[z]}_{\gamma}$ & $-0.0008$ ($-0.0024$, $0.0013$) & $-0.0004$ ($-0.0018$, $0.0011$) \\
\hline
\hline
\multirow{4}{*}{\textbf{\makecell[l]{Grow Factor \\ Variances of Z}}} 
& $\psi^{[z]}_{00}$ & $-0.0150$ ($-0.0230$, $-0.0044$) & $-0.0064$ ($-0.0151$, $0.0004$) \\
& $\psi^{[z]}_{11}$ & $0.0415$ ($0.0146$, $0.0767$) & $0.0068$ ($-0.0034$, $0.0289$) \\
& $\psi^{[z]}_{22}$ & $0.0677$ ($0.0395$, $0.1314$) & $0.0118$ ($-0.0026$, $0.0614$) \\
& $\psi^{[z]}_{\gamma\gamma}$ & --- & $0.1689$ ($-0.2463$, $0.3152$) \\
\hline
\hline
\multirow{4}{*}{\textbf{\makecell[l]{Grow Factor \\ Covariances \\ of Y and Z}}} 
& $\psi^{[yz]}_{00}$\tnote{3} & $-0.0468$ (NA\tnote{2}, NA) & $-0.0136$ (NA, NA) \\
& $\psi^{[yz]}_{11}$\tnote{4} & $0.2200$ (NA, NA) & $0.0599$ (NA, NA) \\
& $\psi^{[yz]}_{22}$\tnote{5} & $0.1966$ (NA, NA) & $0.0465$ (NA, NA) \\
& $\psi^{[yz]}_{\gamma\gamma}$\tnote{6} & --- & $-0.2634$ (NA, NA) \\
\hline
\hline
\end{tabular}
\label{tbl:rBias10_discription}
\begin{tablenotes}
\small
\item[1] {--- indicates that the relative biases are not available from the reduced PBLSGM.}
\item[2] {NA indicates that the bounds of relative bias is not available. The model performance under the conditions with $0$ population value of between-construct correlation is of interest where the relative bias of those correlations would go infinity.} \\
\item[3] {Bias of $\psi^{[yz]}_{00}$: reduced model: $-0.0190$ ($-0.4455$, $0.4793$); full model: $-0.0128$ ($-0.2265$, $0.2425$)} \\
\item[4] {Bias of $\psi^{[yz]}_{11}$: reduced model: $0.0014$ ($-0.0707$, $0.0739$); full model: $0.0016$ ($-0.0363$, $0.0313$)} \\
\item[5] {Bias of $\psi^{[yz]}_{22}$: reduced model: $0.0016$ ($-0.0733$, $0.0768$); full model: $0.0005$ ($-0.0316$, $0.0357$)} \\
\item[6] {Bias of $\psi^{[yz]}_{\gamma\gamma}$: reduced model: ---; full model: $0.0008$ ($-0.0172$, $0.0319$)} \\
\end{tablenotes}
\end{threeparttable}
\end{table}

\begin{table}[!ht]
\centering
\begin{threeparttable}
\setlength{\tabcolsep}{5pt}
\renewcommand{\arraystretch}{0.75}
\caption{Median and range of the Empirical Standard Error of Each Parameter in PBLSGM in the ITPs Framework ($10$ Repeated Measurements)}
\begin{tabular}{p{3.0cm}p{1cm}R{5.5cm}R{5.5cm}}
\hline
\hline
& \textbf{Para.} & \textbf{Reduced PBLSGM} & \textbf{Full PBLSGM} \\
\hline
& & Median (Range) & Median (Range) \\
\hline
\hline
\multirow{4}{*}{\textbf{\makecell[l]{Grow Factor \\ Means of Y}}} 
& $\mu^{[y]}_{\eta_{0}}$ & $0.2947$ ($0.2214$, $0.3709$) & $0.2949$ ($0.2216$, $0.3708$) \\
& $\mu^{[y]}_{\eta_{1}}$ & $0.0640$ ($0.0447$, $0.0849$) & $0.0648$ ($0.0447$, $0.0854$) \\
& $\mu^{[y]}_{\eta_{2}}$ & $0.0616$ ($0.0447$, $0.0806$) & $0.0616$ ($0.0447$, $0.0812$) \\
& $\mu^{[y]}_{\gamma}$ & $0.0381$ ($0.0265$, $0.0583$) & $0.0394$ ($0.0265$, $0.0592$) \\
\hline
\hline
\multirow{4}{*}{\textbf{\makecell[l]{Grow Factor \\ Variances of Y}}} 
& $\psi^{[y]}_{00}$ & $2.0817$ ($1.5473$, $2.6681$) & $2.1182$ ($1.5657$, $2.6967$) \\
& $\psi^{[y]}_{11}$ & $0.0967$ ($0.0678$, $0.1349$) & $0.1049$ ($0.0678$, $0.1466$) \\
& $\psi^{[y]}_{22}$ & $0.0925$ ($0.0671$, $0.1225$) & $0.0946$ ($0.0656$, $0.1265$) \\
& $\psi^{[y]}_{\gamma\gamma}$ & ---\tnote{1} & $0.0510$ ($0.0245$, $0.0735$) \\
\hline
\hline
\multirow{4}{*}{\textbf{\makecell[l]{Grow Factor \\ Means of Z}}} 
& $\mu^{[z]}_{\eta_{0}}$ & $0.2950$ ($0.2186$, $0.3788$) & $0.2950$ ($0.2189$, $0.3788$) \\
& $\mu^{[z]}_{\eta_{1}}$ & $0.0608$ ($0.0447$, $0.0794$) & $0.0612$ ($0.0447$, $0.0794$) \\
& $\mu^{[z]}_{\eta_{2}}$ & $0.0640$ ($0.0447$, $0.0843$) & $0.0640$ ($0.0458$, $0.0849$) \\
& $\mu^{[z]}_{\gamma}$ & $0.0458$ ($0.0265$, $0.0748$) & $0.0510$ ($0.0265$, $0.0812$) \\
\hline
\hline
\multirow{4}{*}{\textbf{\makecell[l]{Grow Factor \\ Variances of Z}}} 
& $\psi^{[z]}_{00}$ & $2.0812$ ($1.5534$, $2.6559$) & $2.1027$ ($1.5754$, $2.6736$) \\
& $\psi^{[z]}_{11}$ & $0.0903$ ($0.0648$, $0.1204$) & $0.0919$ ($0.0640$, $0.1265$) \\
& $\psi^{[z]}_{22}$ & $0.0975$ ($0.0663$, $0.1345$) & $0.1022$ ($0.0671$, $0.1463$) \\
& $\psi^{[z]}_{\gamma\gamma}$ & --- & $0.0781$ ($0.0245$, $0.1308$) \\
\hline
\hline
\multirow{4}{*}{\textbf{\makecell[l]{Grow Factor \\ Covariances \\ of Y and Z}}} 
& $\psi^{[yz]}_{00}$\tnote{2} & $1.5302$ ($1.0928$, $1.9406$) & $1.5500$ ($1.1061$, $1.9741$) \\
& $\psi^{[yz]}_{11}$\tnote{3} & $0.0693$ ($0.0490$, $0.0922$) & $0.0711$ ($0.0480$, $0.0985$) \\
& $\psi^{[yz]}_{22}$\tnote{4} & $0.0689$ ($0.0500$, $0.0959$) & $0.0700$ ($0.0480$, $0.1015$) \\
& $\psi^{[yz]}_{\gamma\gamma}$\tnote{5} & --- & $0.0339$ ($0.0173$, $0.0500$) \\
\hline
\hline
\end{tabular}
\label{tbl:empSE10_discription}
\begin{tablenotes}
\small
\item[1] {--- indicates that the empirical SEs are not available from the reduced PBLSGM.}
\end{tablenotes}
\end{threeparttable}
\end{table}

\begin{table}[!ht]
\centering
\begin{threeparttable}
\setlength{\tabcolsep}{5pt}
\renewcommand{\arraystretch}{0.75}
\caption{Median and range of the Relative RMSE of Each Parameter in PBLSGM in the ITPs Framework ($10$ Repeated Measurements)}
\begin{tabular}{p{3.0cm}p{1cm}R{5.5cm}R{5.5cm}}
\hline
\hline
& \textbf{Para.} & \textbf{Reduced PBLSGM} & \textbf{Full PBLSGM} \\
\hline
& & Median (Range) & Median (Range) \\
\hline
\hline
\multirow{4}{*}{\textbf{\makecell[l]{Grow Factor \\ Means of Y}}} 
& $\mu^{[y]}_{\eta_{0}}$ & $0.0029$ ($0.0022$, $0.0038$) & $0.0029$ ($0.0022$, $0.0038$) \\
& $\mu^{[y]}_{\eta_{1}}$ & $0.0136$ ($0.0091$, $0.0194$) & $0.0136$ ($0.0091$, $0.0195$) \\
& $\mu^{[y]}_{\eta_{2}}$ & $0.0262$ ($0.0175$, $0.0405$) & $0.0262$ ($0.0176$, $0.0407$) \\
& $\mu^{[y]}_{\gamma}$ & $0.0098$ ($0.0059$, $0.0169$) & $0.0100$ ($0.0061$, $0.0171$) \\
\hline
\hline
\multirow{4}{*}{\textbf{\makecell[l]{Grow Factor \\ Variances of Y}}} 
& $\psi^{[y]}_{00}$ & $0.0853$ ($0.0634$, $0.1092$) & $0.0844$ ($0.0626$, $0.1091$) \\
& $\psi^{[y]}_{11}$ & $0.1361$ ($0.0988$, $0.1821$) & $0.1086$ ($0.0682$, $0.1628$) \\
& $\psi^{[y]}_{22}$ & $0.1104$ ($0.0784$, $0.1430$) & $0.0954$ ($0.0659$, $0.1320$) \\
& $\psi^{[y]}_{\gamma\gamma}$ & ---\tnote{1} & $0.5802$ ($0.2728$, $0.8778$) \\
\hline
\hline
\multirow{4}{*}{\textbf{\makecell[l]{Grow Factor \\ Means of Z}}} 
& $\mu^{[z]}_{\eta_{0}}$ & $0.0028$ ($0.0022$, $0.0037$) & $0.0028$ ($0.0022$, $0.0037$) \\
& $\mu^{[z]}_{\eta_{1}}$ & $0.0140$ ($0.0089$, $0.0221$) & $0.0140$ ($0.0089$, $0.0221$) \\
& $\mu^{[z]}_{\eta_{2}}$ & $0.0236$ ($0.0133$, $0.0423$) & $0.0238$ ($0.0134$, $0.0424$) \\
& $\mu^{[z]}_{\gamma}$ & $0.0099$ ($0.0051$, $0.0166$) & $0.0101$ ($0.0049$, $0.0180$) \\
\hline
\hline
\multirow{4}{*}{\textbf{\makecell[l]{Grow Factor \\ Variances of Z}}} 
& $\psi^{[z]}_{00}$ & $0.0839$ ($0.0640$, $0.1067$) & $0.0831$ ($0.0631$, $0.1072$) \\
& $\psi^{[z]}_{11}$ & $0.1048$ ($0.0687$, $0.1388$) & $0.0916$ ($0.0637$, $0.1294$) \\
& $\psi^{[z]}_{22}$ & $0.1210$ ($0.0806$, $0.1872$) & $0.1026$ ($0.0674$, $0.1587$) \\
& $\psi^{[z]}_{\gamma\gamma}$ & --- & $0.8881$ ($0.2763$, $1.4587$) \\
\hline
\hline
\multirow{4}{*}{\textbf{\makecell[l]{Grow Factor \\ Covariances \\ of Y and Z}}} 
& $\psi^{[yz]}_{00}$\tnote{3} & $0.2100$ ($-0.2634$, NA\tnote{2} ) & $0.2088$ ($-0.2642$, NA) \\
& $\psi^{[yz]}_{11}$\tnote{4} & $0.3192$ ($-0.3820$, NA) & $0.2496$ ($-0.3438$, NA) \\
& $\psi^{[yz]}_{22}$\tnote{5} & $0.3162$ ($-0.3846$, NA) & $0.2465$ ($-0.3378$, NA) \\
& $\psi^{[yz]}_{\gamma\gamma}$\tnote{6} & --- & $1.3182$ ($-2.0787$, NA) \\
\hline
\hline
\end{tabular}
\label{tbl:rRMSE10_discription}
\begin{tablenotes}
\small
\item[1] {--- indicates that the relative RMSEs are not available from the reduced PBLSGM.}
\item[2] {NA indicates that the upper bound of the relative RMSE is not available. The model performance under the conditions with $0$ population value of between-construct correlation is of interest where the relative bias of those correlations would go infinity.} \\
\item[3] {RMSE of $\psi^{[yz]}_{00}$: reduced model: $1.5379$ ($1.0940$, $1.9753$); full model: $1.5379$ ($1.1071$, $1.9817$)} \\
\item[4] {RMSE of $\psi^{[yz]}_{11}$: reduced model: $0.0865$ ($0.0489$, $0.1146$); full model: $0.0731$ ($0.0481$, $0.1031$)} \\
\item[5] {RMSE of $\psi^{[yz]}_{22}$: reduced model: $0.0840$ ($0.0501$, $0.1227$); full model: $0.0710$ ($0.0484$, $0.1074$)} \\
\item[6] {RMSE of $\psi^{[yz]}_{\gamma\gamma}$: reduced model: ---; full model: $0.0348$ ($0.0166$, $0.0561$)} \\
\end{tablenotes}
\end{threeparttable}
\end{table}

\begin{table}[!ht]
\centering
\begin{threeparttable}
\setlength{\tabcolsep}{5pt}
\renewcommand{\arraystretch}{0.75}
\caption{Median and range of the Coverage Probability of Each Parameter in PBLSGM in the ITPs Framework ($10$ Repeated Measurements)}
\begin{tabular}{p{3.0cm}p{1cm}R{5.0cm}R{5.0cm}}
\hline
\hline
& \textbf{Para.} & \textbf{Reduced PBLSGM} & \textbf{Full PBLSGM} \\
\hline
& & Median (Range) & Median (Range) \\
\hline
\hline
\multirow{4}{*}{\textbf{\makecell[l]{Grow Factor \\ Means of Y}}} 
& $\mu^{[y]}_{\eta_{0}}$ & $0.9455$ ($0.9200$, $0.9590)$ & $0.9484$ ($0.9153$, $0.9656)$\tnote{1} \\
& $\mu^{[y]}_{\eta_{1}}$ & $0.9550$ ($0.9410$, $0.9710)$ & $0.9468$ ($0.9196$, $0.9626)$ \\
& $\mu^{[y]}_{\eta_{2}}$ & $0.9535$ ($0.9380$, $0.9670)$ & $0.9478$ ($0.9276$, $0.9635)$ \\
& $\mu^{[y]}_{\gamma}$ & $0.9040$ ($0.8560$, $0.9340)$ & $0.9500$ ($0.9216$, $0.9675)$ \\
\hline
\hline
\multirow{4}{*}{\textbf{\makecell[l]{Grow Factor \\ Variances of Y}}} 
& $\psi^{[y]}_{00}$ & $0.9280$ ($0.9110$, $0.9500)$ & $0.9452$ ($0.9254$, $0.9597)$ \\
& $\psi^{[y]}_{11}$ & $0.8865$ ($0.6610$, $0.9490)$ & $0.9496$ ($0.9311$, $0.9655)$ \\
& $\psi^{[y]}_{22}$ & $0.9355$ ($0.8450$, $0.9630)$ & $0.9476$ ($0.9303$, $0.9619)$ \\
& $\psi^{[y]}_{\gamma\gamma}$ & ---\tnote{2} & $0.9772$ ($0.9440$, $0.9954)$ \\
\hline
\hline
\multirow{4}{*}{\textbf{\makecell[l]{Grow Factor \\ Means of Z}}} 
& $\mu^{[z]}_{\eta_{0}}$ & $0.9470$ ($0.9290$, $0.9620)$ & $0.9486$ ($0.9265$, $0.9664)$ \\
& $\mu^{[z]}_{\eta_{1}}$ & $0.9520$ ($0.9360$, $0.9630)$ & $0.9466$ ($0.9244$, $0.9662)$ \\
& $\mu^{[z]}_{\eta_{2}}$ & $0.9540$ ($0.9370$, $0.9670)$ & $0.9492$ ($0.9249$, $0.9662)$ \\
& $\mu^{[z]}_{\gamma}$ & $0.9110$ ($0.8460$, $0.9450)$ & $0.9528$ ($0.9367$, $0.9763)$ \\
\hline
\hline
\multirow{4}{*}{\textbf{\makecell[l]{Grow Factor \\ Variances of Z}}} 
& $\psi^{[z]}_{00}$ & $0.9340$ ($0.9180$, $0.9480)$ & $0.9437$ ($0.9249$, $0.9586)$ \\
& $\psi^{[z]}_{11}$ & $0.9460$ ($0.8570$, $0.9670)$ & $0.9458$ ($0.9148$, $0.9664)$ \\
& $\psi^{[z]}_{22}$ & $0.9285$ ($0.6640$, $0.9610)$ & $0.9492$ ($0.9276$, $0.9662)$ \\
& $\psi^{[z]}_{\gamma\gamma}$ & --- & $0.9712$ ($0.9343$, $0.9889)$ \\
\hline
\hline
\multirow{4}{*}{\textbf{\makecell[l]{Grow Factor \\ Covariances \\ of Y and Z}}} 
& $\psi^{[yz]}_{00}$ & $0.9405$ ($0.9190$, $0.9610)$ & $0.9483$ ($0.9306$, $0.9696)$ \\
& $\psi^{[yz]}_{11}$ & $0.9140$ ($0.7360$, $0.9610)$ & $0.9492$ ($0.9206$, $0.9684)$ \\
& $\psi^{[yz]}_{22}$ & $0.9175$ ($0.7330$, $0.9630)$ & $0.9516$ ($0.9333$, $0.9673)$ \\
& $\psi^{[yz]}_{\gamma\gamma}$ & --- & $0.9678$ ($0.9329$, $0.9919)$ \\
\hline
\hline
\end{tabular}
\label{tbl:CP10_discription}
\begin{tablenotes}
\small
\item[1] {For the full PBLSGM, the reported coverage probabilities have four decimals since we calculated the coverage probabilities only based on the replications with proper solutions.}
\item[2] {--- indicates that the coverage probabilities are not available from the reduced PBLSGM.}
\end{tablenotes}
\end{threeparttable}
\end{table}

\begin{table}[!ht]
\centering
\begin{threeparttable}
\setlength{\tabcolsep}{2pt}
\renewcommand{\arraystretch}{0.75}
\caption{Summary of Model Fit Information For the Univariate Development Model of Reading, Mathematics and Science Ability}
\begin{tabular}{p{5cm}rrrrr}
\hline
\hline
\multicolumn{6}{c}{\textbf{Reading}} \\
\hline
\textbf{Model} & \textbf{-2ll} & \textbf{AIC}& \textbf{BIC}& \textbf{\# of Para.} & \textbf{Residuals} \\
\hline
Linear & $28640.69$ & $28653$ & $28677$ & $6$ & $120.32$ \\
\hline
Quadratic & $26327.72$ & $26348$ & $26388$ & $10$ & $47.74$ \\
\hline
Jenss-Bayley & $26252.45$ & $26274$ & $26318$ & $11$ & $45.76$ \\
\hline
BLSGM (Fixed Knot) & $26144.98$ & $26167$ & $26211$ & $11$ & $44.33$ \\
\hline
BLSGM (Random Knot) & $26048.55$ & $26079$ & $26138$ & $15$ & $42.92$ \\
\hline
\hline
\multicolumn{6}{c}{\textbf{Mathematics}} \\
\hline
\textbf{Model} & \textbf{-2ll} & \textbf{AIC}& \textbf{BIC}& \textbf{\# of Para.} & \textbf{Residuals} \\
\hline
Linear & $27203.38$ & $27215$ & $27239$ & $6$ & $74.58$ \\
\hline
Quadratic & $25050.13$ & $25070$ & $25110$ & $10$ & $31.76$ \\
\hline
Jenss-Bayley & $25058.62$ & $25081$ & $25125$ & $11$ & $31.85$ \\
\hline
BLSGM (Fixed Knot) & $25143.70$ & $25166$ & $25210$ & $11$ & $33.31$ \\
\hline
BLSGM (Random Knot) & $25093.62$ & $25124$ & $25183$ & $15$ & $31.77$ \\
\hline
\hline
\multicolumn{6}{c}{\textbf{Science}} \\
\hline
\textbf{Model} & \textbf{-2ll} & \textbf{AIC}& \textbf{BIC}& \textbf{\# of Para.} & \textbf{Residuals} \\
\hline
Linear & $20847.01$ & $20859$ & $20883$ & $6$ & $23.40$ \\
\hline
Quadratic & $20526.14$ & $20546$ & $20586$ & $10$ & $19.75$ \\
\hline
Jenss-Bayley & $20527.46$ & $20549$ & $20593$ & $11$ & $19.75$ \\
\hline
BLSGM (Fixed Knot) & $20606.73$ & $20629$ & $20673$ & $11$ & $20.63$ \\
\hline
BLSGM (Random Knot) & $20481.91$ & $20512$ & $20572$ & $15$ & $18.70$ \\
\hline
\hline
\end{tabular}
\label{tbl:info_uni}
\end{threeparttable}
\end{table}

\begin{table}[!ht]
\centering
\begin{threeparttable}
\setlength{\tabcolsep}{2pt}
\renewcommand{\arraystretch}{0.75}
\caption{Summary of Model Fit Information For the Joint Development Model of Reading and Mathematics Ability}
\begin{tabular}{lrrrrrrr}
\hline
\hline
\multicolumn{8}{c}{\textbf{Models for Main Analysis (Time Structure: Age in Months)}} \\
\hline
\textbf{Model} & \textbf{-2ll} & \textbf{AIC}& \textbf{BIC}& \textbf{\# of Para.} & \textbf{Reading Res.} & \textbf{Math Res.} & \textbf{Res. Cov.}\\
\hline
Parallel Linear & $53915.82$ & $53950$ & $54018$ & $17$ & $120.36$ & $74.58$ & $59.01$ \\
\hline
Parallel Quadratic & $50658.72$ & $50719$ & $50838$ & $30$ & $48.05$ & $31.96$ & $6.04$ \\
\hline
Parallel Jenss-Bayley & $50588.16$ & $50652$ & $50780$ & $32$ & $46.13$ & $32.04$ & $5.85$ \\
\hline
Reduced PBLSGM & $50586.42$ & $50650$ & $50778$ & $32$ & $44.57$ & $33.43$ & $7.42$ \\
\hline
Full PBLSGM & $50437.79$ & $50532$ & $50719$ & $47$ & $42.93$ & $31.94$ & $6.87$ \\
\hline
\hline
\multicolumn{8}{c}{\textbf{Models for Sensitivity Analysis (Time Structure: Grade in Months)}} \\
\hline
\textbf{Model} & \textbf{-2ll} & \textbf{AIC}& \textbf{BIC}& \textbf{\# of Para.} & \textbf{Reading Res.} & \textbf{Math Res.} & \textbf{Res. Cov.}\\
\hline
Reduced PBLSGM & $50604.65$ & $50669$ & $50796$ & $32$ & $44.07$ & $33.78$ & $8.41$ \\
\hline
Full PBLSGM & $50479.22$ & $50573$ & $50761$ & $47$ & $42.28$ & $31.95$ & $8.51$ \\
\hline
\hline
\end{tabular}
\label{tbl:info_RM}
\end{threeparttable}
\end{table}

\begin{table}[!ht]
\centering
\begin{threeparttable}
\setlength{\tabcolsep}{2pt}
\renewcommand{\arraystretch}{0.75}
\caption{Estimates of Parallel Bilinear Spline Growth Model for Reading and Mathematics Ability}
\begin{tabular}{lrrrrrr}
\hline
\hline
& \multicolumn{2}{c}{\textbf{Reading IRT Scores}} & \multicolumn{2}{c}{\textbf{Math IRT Scores}}& \multicolumn{2}{c}{\textbf{Covariances}} \\
\hline
\textbf{Mean} & Estimate (SE) & P value & Estimate (SE) & P value & Estimate (SE) & P value \\
\hline
\textbf{Intercept$^{1}$} & $42.138$ ($0.753$) & $<0.0001^{\ast}$\tnote{2} & $26.355$ ($0.585$) & $<0.0001^{\ast}$ & $-$\tnote{3} & $-$ \\
\textbf{Slope $1$} & $2.030$ ($0.028$) & $<0.0001^{\ast}$ & $1.787$ ($0.019$) & $<0.0001^{\ast}$ & $-$ & $-$ \\
\textbf{Slope $2$} & $0.678$ ($0.015$) & $<0.0001^{\ast}$ & $0.737$ ($0.017$) & $<0.0001^{\ast}$ & $-$ & $-$ \\
\textbf{Knot} & $94.606$ ($0.362$) & $<0.0001^{\ast}$ & $99.989$ ($0.440$) & $<0.0001^{\ast}$ & $-$ & $-$ \\
\hline
\hline
\textbf{Variance} & Estimate (SE) & P value & Estimate (SE) & P value & Estimate (SE) & P value \\
\hline
\textbf{Intercept} & $172.271$ ($16.628$) & $<0.0001^{\ast}$ & $104.276$ ($10.009$) & $<0.0001^{\ast}$ & $100.360$ ($10.650$) & $<0.0001^{\ast}$ \\
\textbf{Slope $1$} & $0.207$ ($0.023$) & $<0.0001^{\ast}$ & $0.086$ ($0.010$) & $<0.0001^{\ast}$ & $0.085$ ($0.012$) & $<0.0001^{\ast}$ \\
\textbf{Slope $2$} & $0.021$ ($0.006$) & $0.0005^{\ast}$ & $0.033$ ($0.008$) & $<0.0001^{\ast}$ & $0.009$ ($0.005$) & $0.0719$ \\
\textbf{Knot} & $10.632$ ($3.843$) & $0.0057^{\ast}$ & $17.019$ ($5.792$) & $0.0033^{\ast}$ & $7.872$ ($3.227$) & $0.0147^{\ast}$ \\
\hline
\hline
\end{tabular}
\label{tbl:est_RM}
\begin{tablenotes}
\small
\item[1] {Intercept was defined as 60-month old in this case.}\\
\item[2] {$^{\ast}$ indicates statistical significance at $0.05$ level.}\\
\item[3] {$-$ indicates that the metric was not available for the model.}
\end{tablenotes}
\end{threeparttable}
\end{table}

\begin{table}[!ht]
\centering
\begin{threeparttable}
\setlength{\tabcolsep}{2pt}
\renewcommand{\arraystretch}{0.75}
\caption{Summary of Model Fit Information For the Joint Development Model of Mathematics and Science Ability}
\begin{tabular}{lrrrrrrr}
\hline
\hline
\textbf{Model} & \textbf{-2ll} & \textbf{AIC}& \textbf{BIC}& \textbf{\# of Para.} & \textbf{Math Res.} & \textbf{Science Res.} & \textbf{Res. Cov.}\\
\hline
Reduced PBLSGM & $45184.78$ & $45249$ & $45377$ & $32$ & $33.25$ & $19.26$ & $2.44$ \\
\hline
Full PBLSGM & $45123.20$ & $45217$ & $45405$ & $47$ & $31.69$ & $18.80$ & $2.14$ \\
\hline
Mixed PBLSGM & $45132.07$ & $45210$ & $45366$ & $39$ & $31.75$ & $19.28$ & $2.45$ \\
\hline
\hline
\end{tabular}
\label{tbl:info_MS}
\end{threeparttable}
\end{table}

\begin{table}[!ht]
\centering
\begin{threeparttable}
\setlength{\tabcolsep}{5pt}
\renewcommand{\arraystretch}{0.6}
\caption{Estimates of Parallel Bilinear Spline Growth Model for Mathematics and Science Ability}
\begin{tabular}{lrrrrrr}
\hline
\hline
& \multicolumn{2}{c}{\textbf{Math IRT Scores}} & \multicolumn{2}{c}{\textbf{Science IRT Scores}}& \multicolumn{2}{c}{\textbf{Covariances}} \\
\hline
\textbf{Mean} & Estimate (SE) & P value & Estimate (SE) & P value & Estimate (SE) & P value \\
\hline
\textbf{Intercept$^{1}$} & $26.467$ ($0.589$) & $<0.0001^{\ast}$\tnote{2} & $22.055$ ($0.426$) & $<0.0001^{\ast}$ & $-$\tnote{3} & $-$ \\
\textbf{Slope $1$} & $1.777$ ($0.019$) & $<0.0001^{\ast}$ & $0.839$ ($0.015$) & $<0.0001^{\ast}$ & $-$ & $-$ \\
\textbf{Slope $2$} & $0.729$ ($0.017$) & $<0.0001^{\ast}$ & $0.575$ ($0.013$) & $<0.0001^{\ast}$ & $-$ & $-$ \\
\textbf{Knot} & $100.365$ ($0.444$) & $<0.0001^{\ast}$ & $100.081$ ($0.949$) & $<0.0001^{\ast}$ & $-$ & $-$ \\
\hline
\hline
\textbf{Variance} & Estimate (SE) & P value & Estimate (SE) & P value & Estimate (SE) & P value \\
\hline
\textbf{Intercept} & $106.483$ ($10.168$) & $<0.0001^{\ast}$ & $36.673$ ($4.898$) & $<0.0001^{\ast}$ & $40.554$ ($5.468$) & $<0.0001^{\ast}$ \\
\textbf{Slope $1$} & $0.086$ ($0.010$) & $<0.0001^{\ast}$ & $0.046$ ($0.006$) & $<0.0001^{\ast}$ & $0.038$ ($0.006$) & $<0.0001^{\ast}$ \\
\textbf{Slope $2$} & $0.032$ ($0.008$) & $0.0001^{\ast}$& $0.022$ ($0.004$) & $<0.0001^{\ast}$ & $0.007$ ($0.004$) & $0.0801$ \\
\textbf{Knot} & $17.650$ ($6.018$) & $0.0034^{\ast}$ & $-$ & $-$ & $-$ & $-$ \\
\hline
\hline
\end{tabular}
\label{tbl:est_MS}
\begin{tablenotes}
\small
\item[1] {In the joint model of mathematics and science ability, the intercept of mathematics ability is defined as the measurement at $60$-month old, while the intercept of science ability is the measurement half-a-year later.} \\
\item[2] {$^{\ast}$ indicates statistical significance at $0.05$ level.} \\
\item[3] {$-$ indicates that the metric was not available for the model.}
\end{tablenotes}
\end{threeparttable}
\end{table}

\renewcommand\thetable{B.\arabic{table}}
\setcounter{table}{0}
\begin{table}[!ht]
\centering
\begin{threeparttable}
\setlength{\tabcolsep}{5pt}
\renewcommand{\arraystretch}{0.75}
\caption{Median and range of the Relative Bias of Each Parameter in PBLSGM in the ITPs Framework  ($6$ Repeated Measurements)}
\begin{tabular}{p{3.0cm}p{1cm}R{5.5cm}R{5.5cm}}
\hline
\hline
& \textbf{Para.} & \textbf{Reduced PBLSGM} & \textbf{Full PBLSGM} \\
\hline
& & Median (Range) & Median (Range) \\
\hline
\hline
\multirow{4}{*}{\textbf{\makecell[l]{Grow Factor \\ Means of Y}}} 
& $\mu^{[y]}_{\eta_{0}}$ & $0.0001$ ($-0.0003$, $0.0003$) & $0.0001$ ($-0.0003$, $0.0003$) \\
& $\mu^{[y]}_{\eta_{1}}$ & $-0.0025$ ($-0.0040$, $-0.0013$) & $-0.0024$ ($-0.0040$, $-0.0012$) \\
& $\mu^{[y]}_{\eta_{2}}$ & $0.0052$ ($0.0017$, $0.0076$) & $0.0052$ ($0.0016$, $0.0077$) \\
& $\mu^{[y]}_{\gamma}$ & $0.0002$ ($-0.0012$, $0.0019$) & $0.0000$ ($-0.0013$, $0.0013$) \\
\hline
\hline
\multirow{4}{*}{\textbf{\makecell[l]{Grow Factor \\ Variances of Y}}} 
& $\psi^{[y]}_{00}$ & $-0.0153$ ($-0.0241$, $-0.0113$) & $-0.009$ ($-0.0204$, $-0.0041$) \\
& $\psi^{[y]}_{11}$ & $0.1222$ ($0.1147$, $0.1301$) & $0.0768$ ($0.0294$, $0.1134$) \\
& $\psi^{[y]}_{22}$ & $0.1241$ ($0.1117$, $0.1320$) & $0.0778$ ($0.0275$, $0.1057$) \\
& $\psi^{[y]}_{\gamma\gamma}$ & ---\tnote{1} & $-0.4256$ ($-0.5250$, $-0.1253$) \\
\hline
\hline
\multirow{4}{*}{\textbf{\makecell[l]{Grow Factor \\ Means of Z}}} 
& $\mu^{[z]}_{\eta_{0}}$ & $0.0001$ ($-0.0002$, $0.0002$) & $0.0000$ ($-0.0002$, $0.0002$) \\
& $\mu^{[z]}_{\eta_{1}}$ & $-0.0023$ ($-0.0031$, $0.0000$) & $-0.0023$ ($-0.0030$, $0.0002$) \\
& $\mu^{[z]}_{\eta_{2}}$ & $0.0036$ ($0.0014$, $0.0071$) & $0.0036$ ($0.0015$, $0.0070$) \\
& $\mu^{[z]}_{\gamma}$ & $0.0004$ ($-0.0026$, $0.0021$) & $0.0002$ ($-0.0031$, $0.0016$) \\
\hline
\hline
\multirow{4}{*}{\textbf{\makecell[l]{Grow Factor \\ Variances of Z}}} 
& $\psi^{[z]}_{00}$ & $-0.0124$ ($-0.0184$, $-0.0066$) & $-0.0070$ ($-0.0136$, $-0.0013$) \\
& $\psi^{[z]}_{11}$ & $0.0781$ ($0.0675$, $0.1293$) & $0.0580$ ($0.0262$, $0.0997$) \\
& $\psi^{[z]}_{22}$ & $0.0782$ ($0.0627$, $0.1276$) & $0.0574$ ($0.0271$, $0.1038$) \\
& $\psi^{[z]}_{\gamma\gamma}$ & --- & $-0.0163$ ($-0.4717$, $0.1457$) \\
\hline
\hline
\multirow{4}{*}{\textbf{\makecell[l]{Grow Factor \\ Covariances \\ of Y and Z}}} 
& $\psi^{[yz]}_{00}$\tnote{3} & $-0.0353$ (NA\tnote{2}, NA) & $-0.0195$ (NA, NA) \\
& $\psi^{[yz]}_{11}$\tnote{4} & $0.3134$ (NA, NA) & $0.2119$ (NA, NA) \\
& $\psi^{[yz]}_{22}$\tnote{5} & $0.3093$ (NA, NA) & $0.2249$ (NA, NA) \\
& $\psi^{[yz]}_{\gamma\gamma}$\tnote{6} & --- & $-0.3290$ ($-1.4912$, NA) \\
\hline
\hline
\end{tabular}
\label{tbl:rBias6_discription}
\begin{tablenotes}
\small
\item[1] {--- indicates that the relative biases are not available from the reduced PBLSGM.}
\item[2] {NA indicates that the bounds of relative bias is not available. The model performance under the conditions with $0$ population value of between-construct correlation is of interest where the relative bias of those correlations would go infinity.} \\
\item[3] {Bias of $\psi^{[yz]}_{00}$: reduced model: $-0.0185$ ($-0.3682$, $0.3057$); full model: $-0.0106$ ($-0.2919$, $0.2433$)} \\
\item[4] {Bias of $\psi^{[yz]}_{11}$: reduced model: $-0.0002$ ($-0.1180$, $0.1211$) ; full model: $0.0040$ ($-0.0816$, $0.0998$)} \\
\item[5] {Bias of $\psi^{[yz]}_{22}$: reduced model: $-0.0006$ ($-0.1221$, $0.1172$); full model: $0.0038$ ($-0.0850$, $0.0918$)} \\
\item[6] {Bias of $\psi^{[yz]}_{\gamma\gamma}$: reduced model: ---; full model: $0.0114$ ($-0.0108$, $0.0403$)} \\
\end{tablenotes}
\end{threeparttable}
\end{table}

\begin{table}[!ht]
\centering
\begin{threeparttable}
\setlength{\tabcolsep}{5pt}
\renewcommand{\arraystretch}{0.75}
\caption{Median and range of the Empirical Standard Error of Each Parameter in PBLSGM in the ITPs Framework ($6$ Repeated Measurements)}
\begin{tabular}{p{3.0cm}p{1cm}R{5.5cm}R{5.5cm}}
\hline
\hline
& \textbf{Para.} & \textbf{Reduced PBLSGM} & \textbf{Full PBLSGM} \\
\hline
& & Median (Range) & Median (Range) \\
\hline
\hline
\multirow{4}{*}{\textbf{\makecell[l]{Grow Factor \\ Means of Y}}} 
& $\mu^{[y]}_{\eta_{0}}$ & $0.2997$ ($0.2214$, $0.3776$) & $0.2998$ ($0.2216$, $0.3774$) \\
& $\mu^{[y]}_{\eta_{1}}$ & $0.0755$ ($0.0529$, $0.1044$) & $0.0755$ ($0.0529$, $0.1044$) \\
& $\mu^{[y]}_{\eta_{2}}$ & $0.0752$ ($0.0529$, $0.1034$) & $0.0752$ ($0.0529$, $0.1034$) \\
& $\mu^{[y]}_{\gamma}$ & $0.0505$ ($0.0332$, $0.0755$) & $0.0510$ ($0.0332$, $0.0755$) \\
\hline
\hline
\multirow{4}{*}{\textbf{\makecell[l]{Grow Factor \\ Variances of Y}}} 
& $\psi^{[y]}_{00}$ & $2.1228$ ($1.5560$, $2.6410$) & $2.1393$ ($1.5774$, $2.6678$) \\
& $\psi^{[y]}_{11}$ & $0.1290$ ($0.0889$, $0.1865$) & $0.1437$ ($0.1063$, $0.1954$) \\
& $\psi^{[y]}_{22}$ & $0.1292$ ($0.0872$, $0.1819$) & $0.1444$ ($0.1054$, $0.1954$) \\
& $\psi^{[y]}_{\gamma\gamma}$ & ---\tnote{1} & $0.0711$ ($0.0500$, $0.0970$) \\
\hline
\hline
\multirow{4}{*}{\textbf{\makecell[l]{Grow Factor \\ Means of Z}}} 
& $\mu^{[z]}_{\eta_{0}}$ & $0.2974$ ($0.2195$, $0.3728$) & $0.2973$ ($0.2195$, $0.3728$) \\
& $\mu^{[z]}_{\eta_{1}}$ & $0.0745$ ($0.0529$, $0.1025$) & $0.0752$ ($0.0529$, $0.1025$) \\
& $\mu^{[z]}_{\eta_{2}}$ & $0.0742$ ($0.0529$, $0.1020$) & $0.0742$ ($0.0529$, $0.1025$) \\
& $\mu^{[z]}_{\gamma}$ & $0.0644$ ($0.0332$, $0.0975$) & $0.0663$ ($0.0346$, $0.0990$) \\
\hline
\hline
\multirow{4}{*}{\textbf{\makecell[l]{Grow Factor \\ Variances of Z}}} 
& $\psi^{[z]}_{00}$ & $2.1399$ ($1.5399$, $2.7714$) & $2.1494$ ($1.5580$, $2.7784$) \\
& $\psi^{[z]}_{11}$ & $0.1231$ ($0.0872$, $0.1789$) & $0.1366$ ($0.0975$, $0.1913$) \\
& $\psi^{[z]}_{22}$ & $0.1257$ ($0.0849$, $0.1828$) & $0.1384$ ($0.0954$, $0.1967$) \\
& $\psi^{[z]}_{\gamma\gamma}$ & --- & $0.1118$ ($0.0500$, $0.2090$) \\
\hline
\hline
\multirow{4}{*}{\textbf{\makecell[l]{Grow Factor \\ Covariances \\ of Y and Z}}} 
& $\psi^{[yz]}_{00}$\tnote{2} & $1.5354$ ($1.1035$, $2.0258$) & $1.5441$ ($1.1136$, $2.0400$) \\
& $\psi^{[yz]}_{11}$\tnote{3} & $0.0927$ ($0.0632$, $0.1319$) & $0.1027$ ($0.0656$, $0.1493$) \\
& $\psi^{[yz]}_{22}$\tnote{4} & $0.0933$ ($0.0624$, $0.1334$) & $0.1017$ ($0.0656$, $0.1425$) \\
& $\psi^{[yz]}_{\gamma\gamma}$\tnote{5} & --- & $0.0424$ ($0.0245$, $0.0640$) \\
\hline
\hline
\end{tabular}
\label{tbl:empSE6_discription}
\begin{tablenotes}
\small
\item[1] {--- indicates that the empirical SEs are not available from the reduced PBLSGM.}
\end{tablenotes}
\end{threeparttable}
\end{table}

\begin{table}[!ht]
\centering
\begin{threeparttable}
\setlength{\tabcolsep}{5pt}
\renewcommand{\arraystretch}{0.75}
\caption{Median and range of the Relative RMSE of Each Parameter in PBLSGM in the ITPs Framework ($6$ Repeated Measurements)}
\begin{tabular}{p{3.0cm}p{1cm}R{5.5cm}R{5.5cm}}
\hline
\hline
& \textbf{Para.} & \textbf{Reduced PBLSGM} & \textbf{Full PBLSGM} \\
\hline
& & Median (Range) & Median (Range) \\
\hline
\hline
\multirow{4}{*}{\textbf{\makecell[l]{Grow Factor \\ Means of Y}}} 
& $\mu^{[y]}_{\eta_{0}}$ & $0.0030$ ($0.0022$, $0.0039$) & $0.0030$ ($0.0022$, $0.0038$) \\
& $\mu^{[y]}_{\eta_{1}}$ & $0.0160$ ($0.0110$, $0.0231$) & $0.0161$ ($0.0110$, $0.0231$) \\
& $\mu^{[y]}_{\eta_{2}}$ & $0.0328$ ($0.0209$, $0.0507$) & $0.0330$ ($0.0209$, $0.0508$) \\
& $\mu^{[y]}_{\gamma}$ & $0.0202$ ($0.0131$, $0.0301$) & $0.0204$ ($0.0133$, $0.0303$) \\
\hline
\hline
\multirow{4}{*}{\textbf{\makecell[l]{Grow Factor \\ Variances of Y}}} 
& $\psi^{[y]}_{00}$ & $0.0849$ ($0.0640$, $0.1072$) & $0.0847$ ($0.0635$, $0.1069$) \\
& $\psi^{[y]}_{11}$ & $0.1776$ ($0.1543$, $0.2211$) & $0.1628$ ($0.1114$, $0.2199$) \\
& $\psi^{[y]}_{22}$ & $0.1784$ ($0.1517$, $0.2216$) & $0.1620$ ($0.1106$, $0.2196$) \\
& $\psi^{[y]}_{\gamma\gamma}$ & ---\tnote{1} & $0.8842$ ($0.5676$, $1.1596$) \\
\hline
\hline
\multirow{4}{*}{\textbf{\makecell[l]{Grow Factor \\ Means of Z}}} 
& $\mu^{[z]}_{\eta_{0}}$ & $0.0030$ ($0.0022$, $0.0037$) & $0.0030$ ($0.0022$, $0.0037$) \\
& $\mu^{[z]}_{\eta_{1}}$ & $0.0172$ ($0.0107$, $0.0275$) & $0.0173$ ($0.0108$, $0.0275$) \\
& $\mu^{[z]}_{\eta_{2}}$ & $0.0282$ ($0.0156$, $0.0491$) & $0.0285$ ($0.0158$, $0.0494$) \\
& $\mu^{[z]}_{\gamma}$ & $0.0256$ ($0.0135$, $0.0390$) & $0.0266$ ($0.0136$, $0.0398$) \\
\hline
\hline
\multirow{4}{*}{\textbf{\makecell[l]{Grow Factor \\ Variances of Z}}} 
& $\psi^{[z]}_{00}$ & $0.0850$ ($0.0624$, $0.1114$) & $0.0850$ ($0.0625$, $0.1114$) \\
& $\psi^{[z]}_{11}$ & $0.1571$ ($0.1160$, $0.2164$) & $0.1522$ ($0.1028$, $0.2155$) \\
& $\psi^{[z]}_{22}$ & $0.1560$ ($0.1116$, $0.2161$) & $0.1516$ ($0.0993$, $0.2185$) \\
& $\psi^{[z]}_{\gamma\gamma}$ & --- & $1.2594$ ($0.5727$, $2.3268$) \\
\hline
\hline
\multirow{4}{*}{\textbf{\makecell[l]{Grow Factor \\ Covariances \\ of Y and Z}}} 
& $\psi^{[yz]}_{00}$\tnote{3} & $0.2070$ ($-0.2659$, NA\tnote{2}) & $0.2066$ ($-0.2664$, NA) \\
& $\psi^{[yz]}_{11}$\tnote{4} & $0.4696$ ($-0.5896$, NA) & $0.4210$ ($-0.5666$, NA) \\
& $\psi^{[yz]}_{22}$\tnote{5} & $0.4591$ ($-0.5900$, NA) & $0.4172$ ($-0.5440$, NA) \\
& $\psi^{[yz]}_{\gamma\gamma}$\tnote{6} & --- & $1.7354$ ($-2.7123$, NA) \\
\hline
\hline
\end{tabular}
\label{tbl:rRMSE6_discription}
\begin{tablenotes}
\small
\item[1] {--- indicates that the relative RMSEs are not available from the reduced PBLSGM.}
\item[2] {NA indicates that the upper bound of relative RMSE is not available. The model performance under the conditions with $0$ population value of between-construct correlation is of interest where the relative bias of those correlations would go infinity.} \\
\item[3] {RMSE of $\psi^{[yz]}_{00}$: reduced model: $1.5244$ ($1.1036$, $2.0580$); full model: $1.5270$ ($1.1136$, $2.0598$)} \\
\item[4] {RMSE of $\psi^{[yz]}_{11}$: reduced model: $0.1259$ ($0.0633$, $0.1783$); full model: $0.1156$ ($0.0656$, $0.1700$)} \\
\item[5] {RMSE of $\psi^{[yz]}_{22}$: reduced model: $0.1292$ ($0.0624$, $0.1770$); full model: $0.1202$ ($0.0657$, $0.1693$)} \\
\item[6] {RMSE of $\psi^{[yz]}_{\gamma\gamma}$: reduced model: ---; full model: $0.0468$ ($0.0276$, $0.0732$)} \\
\end{tablenotes}
\end{threeparttable}
\end{table}

\begin{table}[!ht]
\centering
\begin{threeparttable}
\setlength{\tabcolsep}{5pt}
\renewcommand{\arraystretch}{0.75}
\caption{Median and range of the Coverage Probability of Each Parameter in PBLSGM in the ITPs Framework ($6$ Repeated Measurements)}
\begin{tabular}{p{3.0cm}p{1cm}R{5.0cm}R{5.0cm}}
\hline
\hline
& \textbf{Para.} & \textbf{Reduced PBLSGM} & \textbf{Full PBLSGM} \\
\hline
& & Median (Range) & Median (Range) \\
\hline
\hline
\multirow{4}{*}{\textbf{\makecell[l]{Grow Factor \\ Means of Y}}} 
& $\mu^{[y]}_{\eta_{0}}$ & $0.9480$ ($0.9280$, $0.9580$) & $0.9494$ ($0.9116$, $0.9622$)\tnote{1} \\
& $\mu^{[y]}_{\eta_{1}}$ & $0.9535$ ($0.9400$, $0.9700$) & $0.9464$ ($0.9112$, $0.9862$) \\
& $\mu^{[y]}_{\eta_{2}}$ & $0.9540$ ($0.9320$, $0.9640$) & $0.9480$ ($0.9249$, $0.9596$) \\
& $\mu^{[y]}_{\gamma}$ & $0.9305$ ($0.8970$, $0.9500$) & $0.9523$ ($0.9291$, $0.9842$) \\
\hline
\hline
\multirow{4}{*}{\textbf{\makecell[l]{Grow Factor \\ Variances of Y}}} 
& $\psi^{[y]}_{00}$ & $0.9330$ ($0.9230$, $0.9500$) & $0.9449$ ($0.9215$, $0.9679$) \\
& $\psi^{[y]}_{11}$ & $0.8805$ ($0.7380$, $0.9490$) & $0.9596$ ($0.9375$, $0.9759$) \\
& $\psi^{[y]}_{22}$ & $0.8740$ ($0.7500$, $0.9400$) & $0.9557$ ($0.9159$, $0.9731$) \\
& $\psi^{[y]}_{\gamma\gamma}$ & ---\tnote{2} & $0.9848$ ($0.9683$, $0.9953$) \\
\hline
\hline
\multirow{4}{*}{\textbf{\makecell[l]{Grow Factor \\ Means of Z}}} 
& $\mu^{[z]}_{\eta_{0}}$ & $0.9480$ ($0.9280$, $0.9610$) & $0.9504$ ($0.9229$, $0.9676$) \\
& $\mu^{[z]}_{\eta_{1}}$ & $0.9535$ ($0.9280$, $0.9670$) & $0.9532$ ($0.9186$, $0.9653$) \\
& $\mu^{[z]}_{\eta_{2}}$ & $0.9530$ ($0.9380$, $0.9700$) & $0.9489$ ($0.9241$, $0.9641$) \\
& $\mu^{[z]}_{\gamma}$ & $0.9345$ ($0.9020$, $0.9630$) & $0.9572$ ($0.9209$, $0.9966$) \\
\hline
\hline
\multirow{4}{*}{\textbf{\makecell[l]{Grow Factor \\ Variances of Z}}} 
& $\psi^{[z]}_{00}$ & $0.9375$ ($0.9190$, $0.9470$) & $0.9476$ ($0.9186$, $0.9609$) \\
& $\psi^{[z]}_{11}$ & $0.9265$ ($0.7720$, $0.9650$) & $0.9604$ ($0.9327$, $0.9802$) \\
& $\psi^{[z]}_{22}$ & $0.9155$ ($0.7480$, $0.9580$) & $0.9580$ ($0.9437$, $0.9728$) \\
& $\psi^{[z]}_{\gamma\gamma}$ & --- & $0.9732$ ($0.9544$, $0.9943$) \\
\hline
\hline
\multirow{4}{*}{\textbf{\makecell[l]{Grow Factor \\ Covariances \\ of Y and Z}}} 
& $\psi^{[yz]}_{00}$ & $0.9395$ ($0.9240$, $0.9620$) & $0.9464$ ($0.9249$, $0.9683$) \\
& $\psi^{[yz]}_{11}$ & $0.8735$ ($0.6190$, $0.9660$) & $0.9565$ ($0.9252$, $0.9750$) \\
& $\psi^{[yz]}_{22}$ & $0.8885$ ($0.5970$, $0.9580$) & $0.9556$ ($0.9309$, $0.9829$) \\
& $\psi^{[yz]}_{\gamma\gamma}$ & --- & $0.9784$ ($0.9596$, $0.9960$) \\
\hline
\hline
\end{tabular}
\label{tbl:CP6_discription}
\begin{tablenotes}
\small
\item[1] {For the full PBLSGM, the reported coverage probabilities have four decimals since we calculated the coverage probabilities only based on the replications with proper solutions.}
\item[2] {--- indicates that the coverage probabilities are not available from the reduced PBLSGM.}
\end{tablenotes}
\end{threeparttable}
\end{table}

\FloatBarrier
\newpage

\renewcommand\thefigure{\arabic{figure}}
\setcounter{figure}{0}
\begin{figure}[!ht]
\centering
\includegraphics[width=0.6\textwidth]{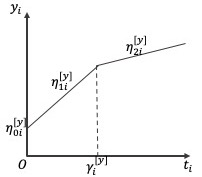}
\caption{Within-individual Change over Time with Bilinear Spline Functional Form}
\label{fig:knot}
\end{figure}

\begin{figure}[!ht]
\begin{subfigure}{.50\textwidth}
\centering
\includegraphics[width=1.0\linewidth]{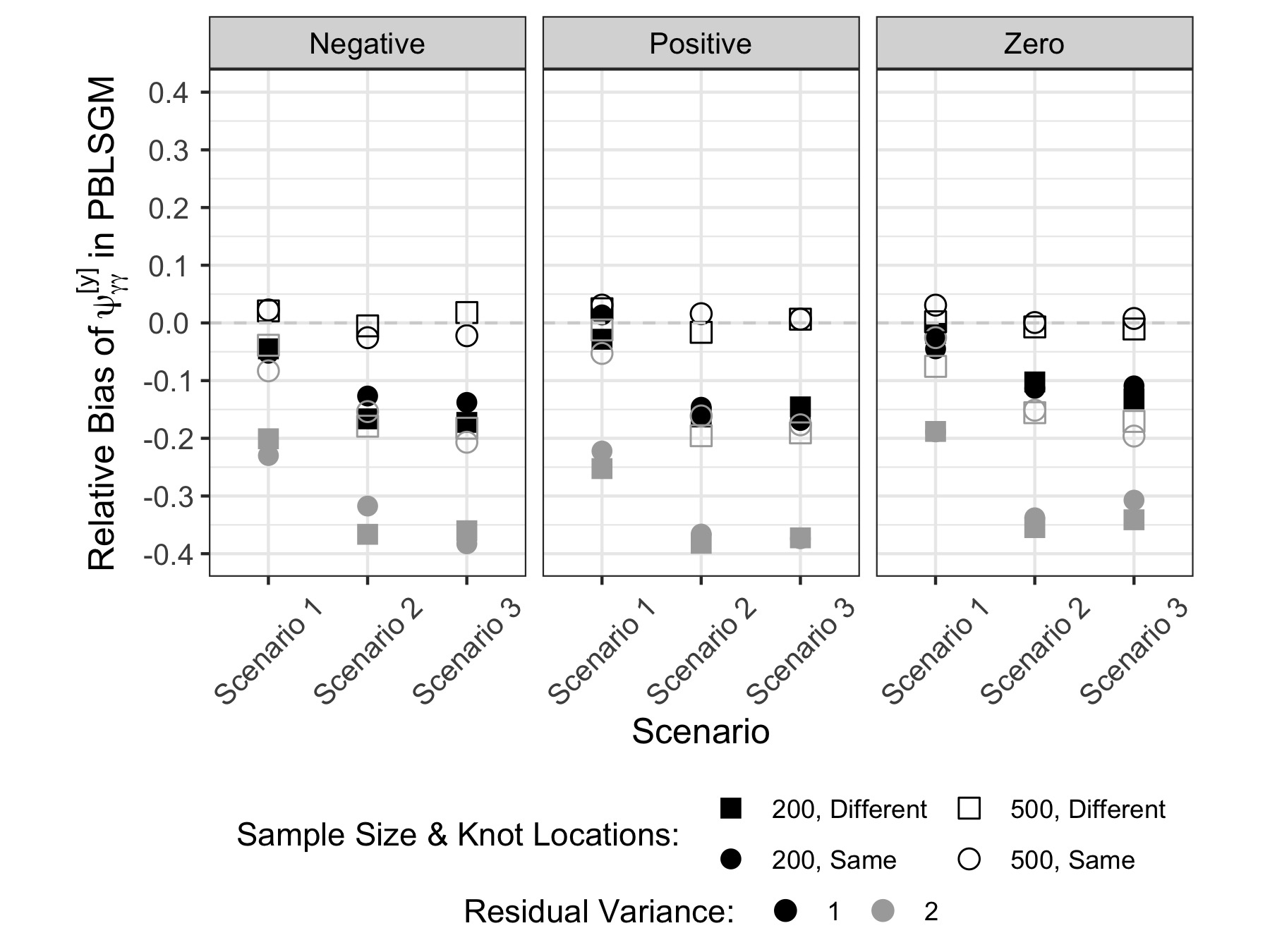}
\caption{Relative Bias of $\psi_{\gamma\gamma}^{[y]}$}
\label{fig:rBiasY}
\end{subfigure}%
\begin{subfigure}{.50\textwidth}
\centering
\includegraphics[width=1.0\linewidth]{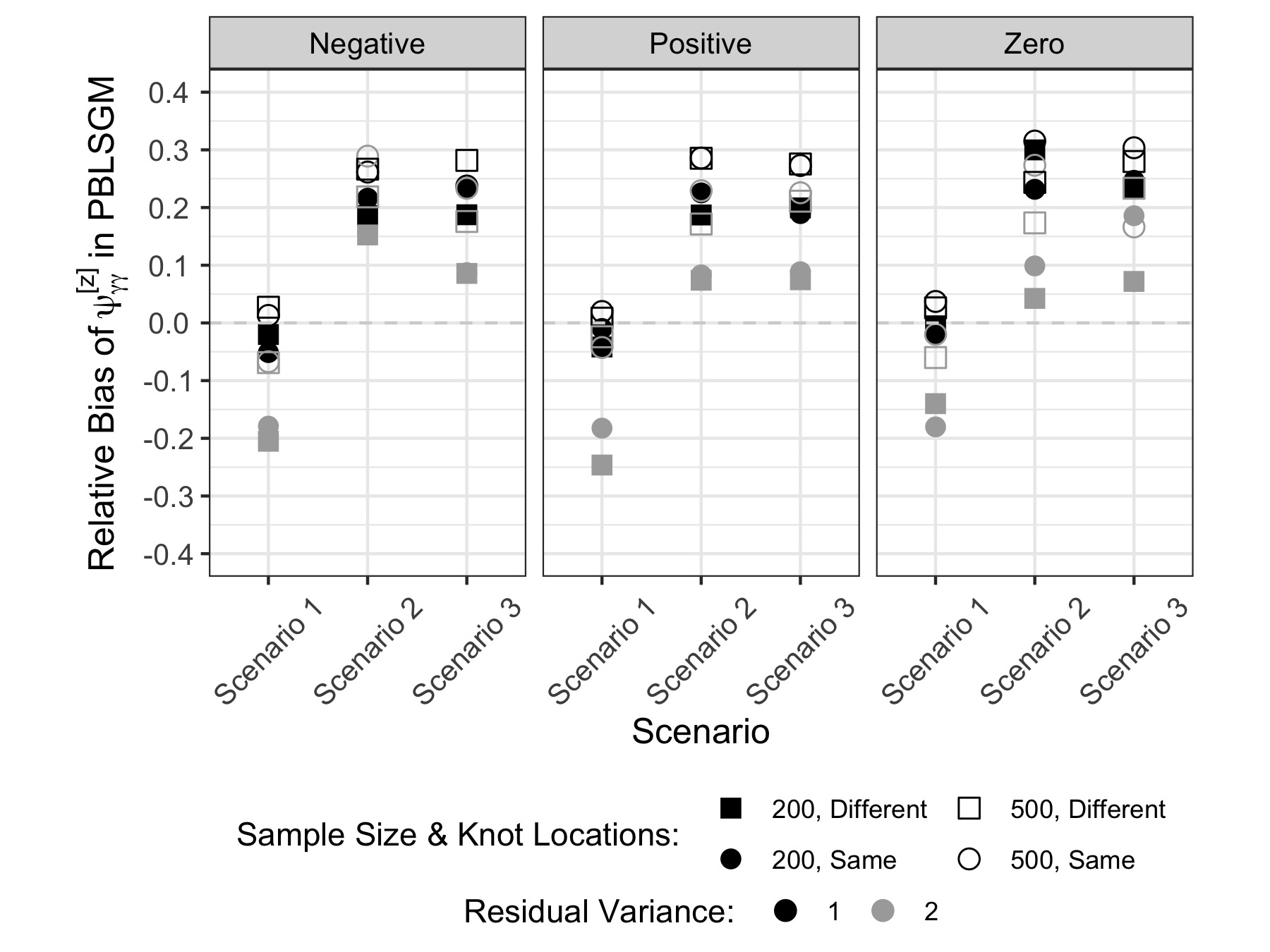}
\caption{Relative Bias of $\psi_{\gamma\gamma}^{[z]}$}
\label{fig:rBiasZ}
\end{subfigure}
\caption{Relative Bias of Knot Variances under Conditions with $10$ Repeated Measures}
\label{fig:rBias}
\end{figure}

\begin{figure}[!ht]
\centering
\includegraphics[width=1.0\textwidth]{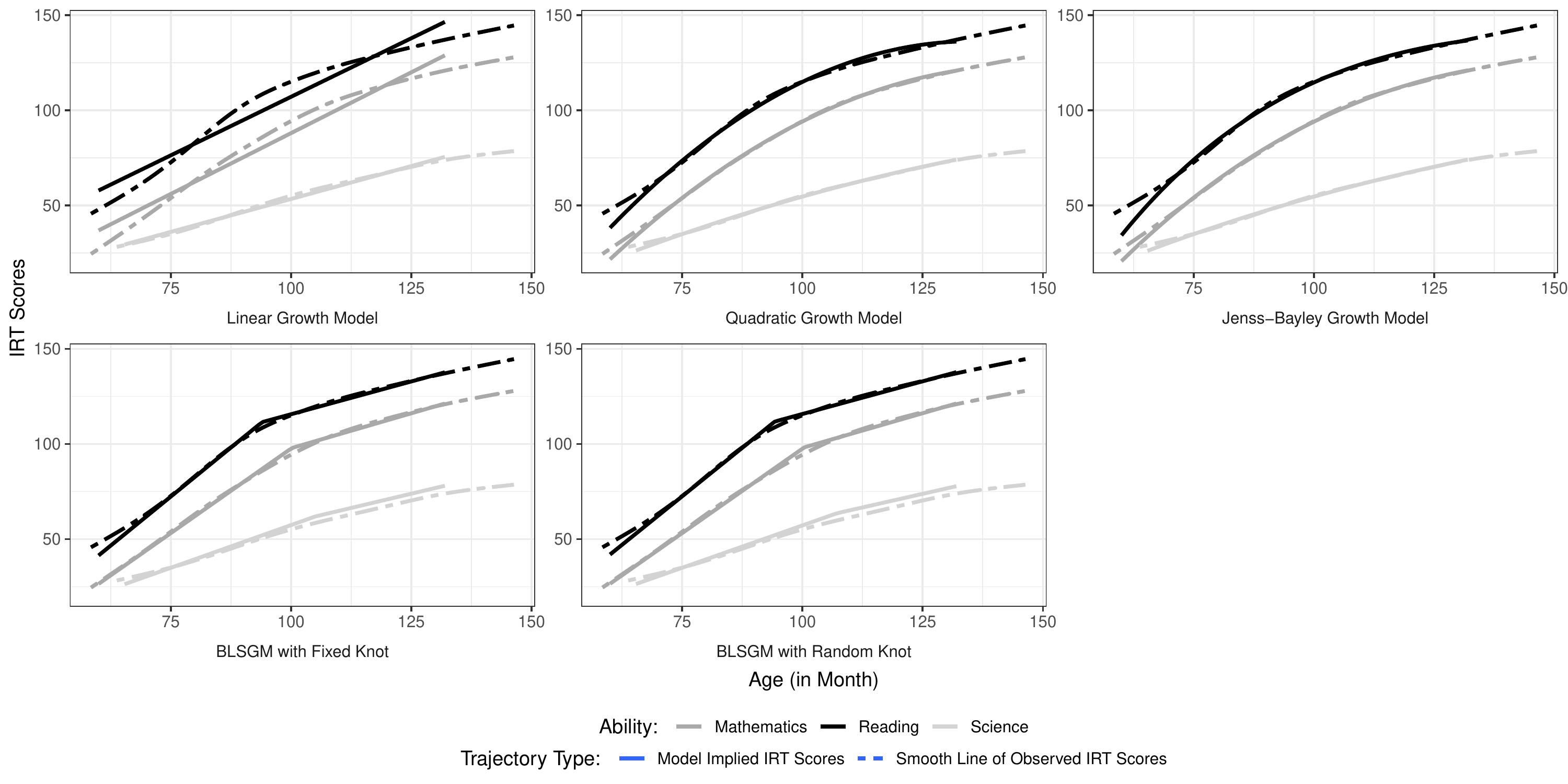}
\caption{Model Implied Trajectory and Smooth Line of Univariate Repeated Outcome}
\label{fig:all_traj}
\end{figure}

\begin{figure}[!ht]
\centering
\includegraphics[width=1.0\textwidth]{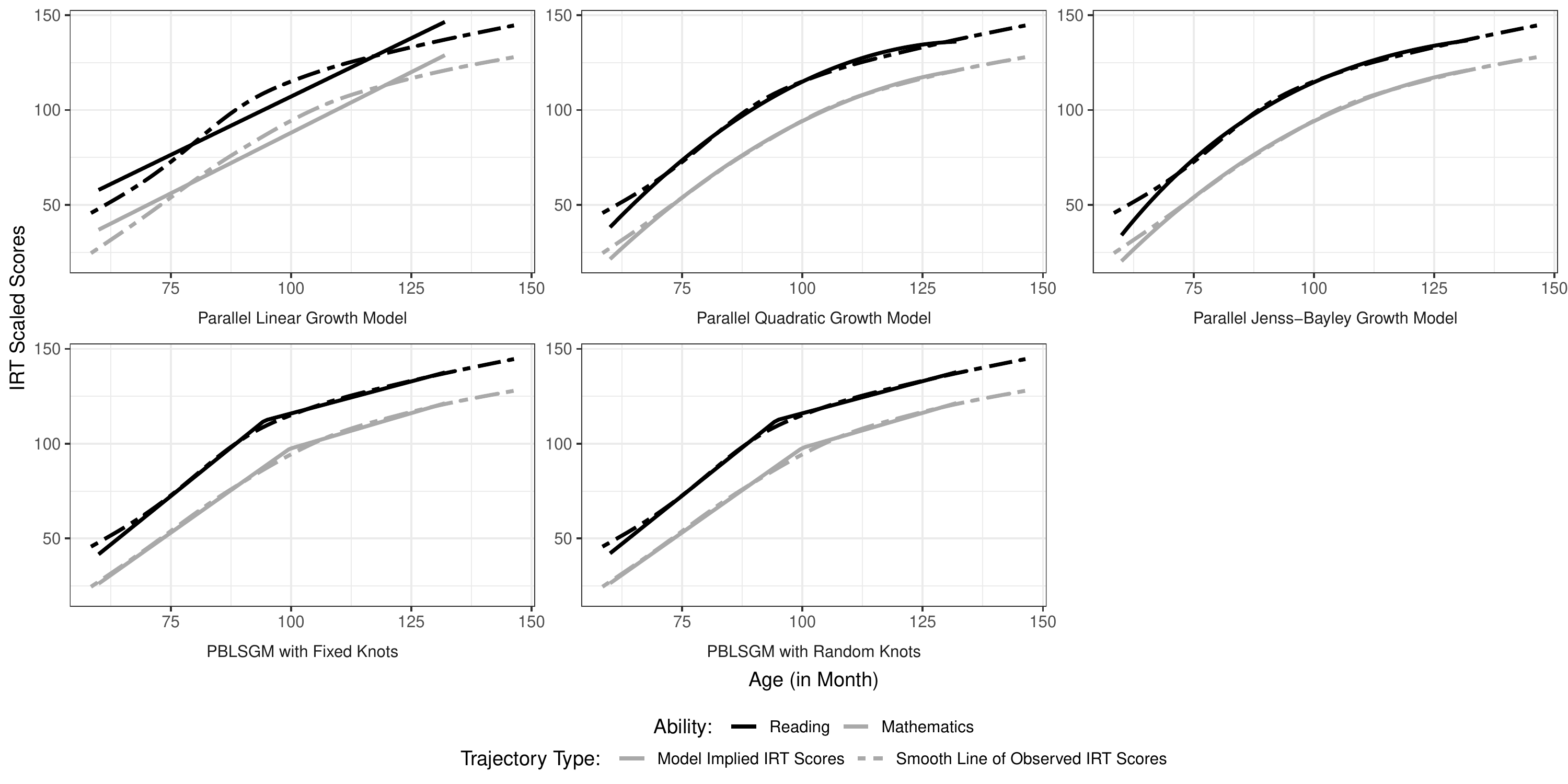}
\caption{Model Implied Trajectory and Smooth Line of Bivariate Outcome: Reading and Mathematics}
\label{fig:RM_traj}
\end{figure}

\renewcommand\thefigure{A.\arabic{figure}}
\setcounter{figure}{0}

\begin{figure}[!ht]
\centering
\includegraphics[width=0.8\textwidth]{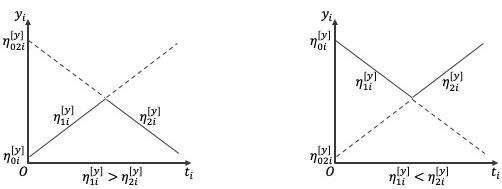}
\caption{The Two Forms of the Bilinear Spline (Linear-Linear Piecewise)}
\label{fig:proj1_2cases}
\end{figure}

\end{document}